\documentclass[10pt,preprint]{aastex}

\def\be{\begin{equation}}
\def\ee{\end{equation}}
\def\d{{\rm d}}

\def\acc{_{\rm acc}}

\def\bh{_{\rm BH}}
\def\bhpr{_{{\rm BH},0}}
\def\bhti{_{{\rm BH},i}}
\def\bhtI{_{{\rm BH,I}}}
\def\bol{_{\rm bol}}

\def\c{_{\rm c}}
\def\D{_{\rm D}}
\def\cT{{\cal T}}
\def\cR{{\cal R}}
\def\e{_{\rm e}}
\def\early{^{\rm early}}
\def\Edd{_{\rm Edd}}
\def\I{_{\rm I}}
\def\late{^{\rm late}}
\def\life{_{\rm life}}

\def\LBQSOopt{_{L_B,{\rm QSO}}^{\rm opt}}
\def\Llocal{_{L,{\rm local}}}
\def\LnuB{{L_{\nu{_B}}}}
\def\LQSO{_{L,{\rm QSO}}}
\def\MBlocal{_{M_B,{\rm local}}}
\def\MBpr{{(M_B,t_0)}}
\def\MBQSO{_{M_B,{\rm QSO}}}
\def\MBQSOopt{_{M_B,{\rm QSO}}^{\rm opt}}

\def\opt{^{\rm opt}}
\def\scatter{{\Delta_{\log M\bhpr}}}
\def\Sp{_{\rm Sp}}

\def\dex{{\rm\,dex}}

\def\kms{{\rm\,km\,s^{-1}}}

\def\msun{{\rm\,M_\odot}}
\def\mag{{\rm\,mag}}
\def\Mpc{{\rm\,Mpc}}
\def\yr{{\rm\,yr}}

\begin{document}
\title{Constraints on QSO models from a relation between the QSO luminosity
function and the local black hole mass function}
\author{Qingjuan Yu$^{1,}$\footnotemark[3] and Youjun Lu$^{1,2,}$\footnotemark[3]}
\affil{$^1$Canadian Institute for Theoretical Astrophysics, 60 St.\ George
Street, Toronto, Ontario M5S 3H8, Canada \\ $^2$ Center for Astrophysics,
University of Science and Technology of China, 96 Jinzhai Road, Hefei,
Anhui 230026, P.\ R.\ China}
\footnotetext[3]{Current address: Astronomy Department, University of
California at Berkeley, Berkeley, CA 94720, USA;\\
Email: yqj@astro.berkeley.edu (QY); lyj@astro.berkeley.edu (YL)}
\begin{abstract}
\noindent
QSOs are believed to be powered by accretion onto massive black holes (BHs).
In this paper, assuming that each central BH in nearby galaxies has
experienced the QSO phase and ignoring BH mergers, we establish a relation
between the QSO luminosity function (LF) and the local BH mass function (MF).
The QSOLF is jointly
controlled by the luminosity evolution of individual QSOs and the triggering
history of the accretion onto seed BHs. By comparing the time integral of the
QSOLF with that inferred from local BHs, we separate the effect of the
luminosity evolution of individual QSOs from the effect of the triggering
history. Assuming that the nuclear luminosity evolution includes two
phases (first increasing at the Eddington luminosity with growth of BHs
and then declining), we
find that observations are generally consistent with the expected relation
between the QSOLF and the local BHMF and obtain the following constraints on
QSO models and BH growth:
(i) The QSO mass-to-energy efficiency $\epsilon$ should be $\ga 0.1$.
(ii) The lifetime (defined directly through the luminosity evolution of
individual QSOs here) should be $\ga 4\times10^7\yr$. The
characteristic declining timescale in the second phase should be significantly
shorter than the Salpeter timescale $\tau\Sp$, and BH growth should not be
dominated by the second phase.
(iii) The ratio of obscured QSOs/AGNs to optically bright
QSOs should be not larger than 7 at $M_B\sim -23$ and 3 at $M_B\sim -26$ if
$\epsilon=0.31$, and not larger than 1 at $M_B\sim -23$
and negligible at $M_B\sim -26$ if $\epsilon=0.1$.
(iv) It is unlikely that most QSOs are accreting at super-Eddington
luminosities.
We point out that it is hard to accurately estimate the value of the QSO lifetime
estimated from the QSOLF and/or the local BHMF, if it is longer than a certain
value (e.g.,\ $\sim4\tau\Sp$ in this study).
We discuss the importance of accurate measurements of the intrinsic
scatter in the BH mass and velocity dispersion relation of local
galaxies and the scatter in the bolometric correction of QSOs.
We also discuss some possible applications of the work in this paper,
such as to the study of the demography of QSOs and the
demography of normal galaxies at intermediate redshift.

\end{abstract}
\keywords{black hole physics -- galaxies: active -- galaxies: evolution --
galaxies: luminosity function, mass function --
galaxies: nuclei -- quasars: general}
\maketitle

\section{Introduction}\label{sec:intro}

\noindent
The exploration of the relation of QSOs with massive black holes (BHs) in
nearby galaxies has been of considerable interests since the discovery of QSOs.
On the one hand, QSOs are suggested to be powered by gas accretion onto
massive BHs \citep{S64,ZN64}.
This (currently widely accepted) model suggests that a population of massive
BHs as ``dead'' QSOs exist in nearby galactic centers \citep{Lynden69}.
Furthermore, the total mass density of these remnant BHs and the typical BH mass in
nearby galaxies can be inferred from the total energy density radiated in
photons by QSOs (\citealt{S82}; see also \citealt{R84}).
These simple and elegant arguments motivate the search for
BHs in nearby galaxies (e.g.,\ \citealt{KR95,Magorrian98,Gebhardt03,K03}).
On the other hand, as a result of the endeavor in the past two decades,
not only has the existence of the massive dark objects (which are presumably
BHs here) in most nearby galactic centers been confirmed, but also dramatic
progress on their demography has been recently achieved
(e.g.,\ \citealt{KG01,Tremaine02} and references therein).
Comparisons between the properties of local BHs and those inferred from
the QSO model may shed new light on our understanding of the BH growth,
the accretion physics, the mechanisms to trigger and quench nuclear activities,
the formation and evolution of galaxies
etc.\ (e.g.\ \citealt{CP89,SB92,FI99,Salucci99,Elvis02,MS02,YT02,Fab03}).

The comparison between the local BH mass density and the local energy density
in QSO photons has suggested that local BHs acquire most of their mass through
accretion during the QSO/AGN phase \citep{YT02,AR02,Fab03}.
The mass distribution of local BHs as the remnant of nuclear activities is
therefore controlled by the triggering history of the accretion onto seed BHs
and the luminosity evolution of individual triggered nuclei with a 
mass-to-energy conversion efficiency $\epsilon$
(i.e., the growth of individual BHs due to accretion).
The triggering rate is usually believed to be related to the formation and
evolution of galaxies and is a function of cosmic time,
and the luminosity evolution of individual nuclei is believed to
contain information on the accretion process in the vicinity of the BH and 
is a function of the physical time spent since the triggering of the accretion
onto seed BHs.
The triggering history and the luminosity evolution jointly control the QSO luminosity function (LF)
as a function of luminosity and redshift;
however, each cannot be uniquely determined from the QSOLF itself,
partly as a result of the mixing of their effects on the QSOLF.
In this paper, by using the local BH mass function (MF) as an additional
constraint on QSO models and establishing a new relation between the
QSOLF and the local BHMF, the luminosity evolution of individual QSOs is
isolated from the triggering history of the QSO population
in their effects on the QSOLF,
and then we use observations to test QSO models and provide constraints
on the QSO luminosity evolution and BH growth.

A relation between the QSOLF and the local BHMF has been achieved by
\citet{YT02} under the assumptions that the seed BH mass is negligible and
the luminosities of QSOs are only an increasing function of their 
central BH mass. In the study of this paper, we relax these two assumptions.
In addition, the relation in \citet{YT02} includes the effect of BH mergers.
However, BH mergers are ignored in the relation established in this paper for
the following reasons:
(i) BH mergers are not shown to play a significant role or not necessarily
required at least for growth of high-mass ($\ga 10^8\msun$) BHs,
if $\epsilon\simeq 0.1$ \citep{YT02};
(ii) currently, the BH merger process and rate are very uncertain;
and (iii) comparison of observations with the expectation obtained by
ignoring BH mergers may also provide considerable insights in the role
of BH mergers. The detailed difference between these two relations
will be further discussed in this paper (see \S~\ref{sec:degeneracy}
and \ref{sec:parameters}).

This paper is organized as follows.
In \S~\ref{sec:relation}, by studying the continuity equation for the BH mass
and nuclear luminosity distribution, we establish the relation between the
QSOLF and the local BHMF. The triggering history of the accretion onto seed
BHs is (implicitly) considered in the continuity equation
but circumvented in the relation between the QSOLF and local BHMF.
Only the luminosity evolution of individual QSOs is incorporated in this
relation.
In \S~\ref{sec:obs} we obtain the local BHMF by using the
Sloan Digital Sky Survey (SDSS) observation results on both early-type
and late-type galaxies and the empirical relations on the demography of
galaxies and BHs (the BH mass and velocity dispersion relation,
the Tully-Fisher relation, etc.).
We also review the QSOLF obtained from large optical surveys.
In \S~\ref{sec:comparison} we combine the observations with the relation
obtained in \S~\ref{sec:relation} to provide accurate observational constraints
on the luminosity evolution of individual QSOs and BH growth,
such as the QSO lifetime, the mass-to-energy conversion efficiency,
the role of obscuration in BH growth, etc.
The results are discussed in \S~\ref{sec:discussion}.
In \S~\ref{sec:application} we further discuss some possible applications
of the framework established in \S~\ref{sec:relation}, such as to the study of
the demography of QSOs and the demography of normal galaxies at
intermediate redshift.
Finally, our conclusions are summarized in \S~\ref{sec:conclusion}.

In this paper we set the Hubble constant as $H_0=100h\kms\Mpc^{-1}$,
and if not otherwise specified, the cosmological model used is
$(\Omega_{\rm M},\Omega_{\Lambda},h)=(0.3,0.7,0.65)$ \citep{Wang00}.

\section{The expected relation between the QSO luminosity function and the local
BH mass function}\label{sec:relation}

\subsection{The continuity equation}\label{sec:cont}
\noindent
In this subsection we describe the evolution of the BH distribution by a
continuity equation, which will be used to establish the relation between the
QSOLF and the local BHMF.

We define ${\cal N}(t_i,M\bhpr,L,t)$ ($t\ge t_i$) so that
${\cal N}(t_i,M\bhpr,L,t)\d t_i\d M\bhpr\d L\d t$ is the comoving
number density of local BHs with such properties:
the nuclear activity due to accretion onto their seed BHs was triggered
during cosmic time $t_i\rightarrow t_i+\d t_i$,
their nuclear luminosities were in the range $L\rightarrow L+\d L$ at cosmic
time $t$,
and these BHs have mass in the range $M\bhpr\rightarrow M\bhpr+\d M\bhpr$
at present time $t_0$.
We assume that the change rate of the nuclear luminosity $\dot L$ is a
function only of $(t_i,M\bhpr,L,t)$.
Ignoring BH mergers, we can use the following continuity equation to
describe the evolution of ${\cal N}(t_i,M\bhpr,L,t)$:
\be
\frac{\partial {\cal N}(t_i,M\bhpr,L,t)}{\partial t}+
\frac{\partial[\dot L(t_i,M\bhpr,L,t){\cal N}(t_i,M\bhpr,L,t)]}{\partial L}=0.
\label{eq:Euler}
\ee
We define the nuclear LF $n_L(L,t)$ as follows:
\be
n_L(L,t)\equiv\int_0^\infty\d M\bhpr\int_0^t\d t_i~{\cal N}(t_i,M\bhpr,L,t)
\label{eq:nL}
\ee
so that $n_L(L,t)\d L$ is the number density of the local BHs that had nuclear
luminosities in the range $L\rightarrow L+\d L$ at cosmic time $t$.

Below we illustrate the relation of equation (\ref{eq:Euler}) with some
continuity equations of the QSOLF or BHMF in the literature
(e.g., \citealt{CP90,SB92}).
By integrating equation (\ref{eq:Euler}) over $M\bhpr$ from 0 to $\infty$ and
over $t_i$ from 0 to $t$, we may obtain the evolution of $n_L(L,t)$ as follows:
\be
\frac{\partial{n_L(L,t)}}{\partial t}+
\frac{\partial{[\langle \dot L\rangle n_L(L,t)]}}{\partial L}=S(L,t),
\label{eq:contS}
\ee
where $\langle\dot L\rangle(L,t)$ is the mean change rate of $L$ defined by
\be
\langle \dot L\rangle\equiv\frac{\int_0^\infty\d M\bhpr\int_0^t\d t_i
~\dot L(t_i,M\bhpr,L,t){\cal N}(t_i,M\bhpr,L,t)}{n_L(L,t)},
\ee
and $S(L,t)$ is the source function defined by
\be
S(L,t)\equiv \int_0^\infty {\cal N}(t_i,M\bhpr,L,t)|_{t_i=t}\d M\bhpr
\ee
and describing the triggering rate of nuclear activities of seed BHs.
If $\dot L(t_i,M\bhpr,L,t)$ is a function only of $L$ and $t$ and
$n_L(L,t)$ is replaced by the QSOLF,
equation (\ref{eq:contS}) will be identical to equation (1) in \citet{Cavaliere71} or equation (3) in \citet{CP90}.
If we replace $L$ in equation (\ref{eq:contS}) with BH mass of the progenitors of local BHs, equation
(\ref{eq:contS}) will look the same as equation (8) in \citet{SB92}
(which describes the evolution of the BHMF).

\subsection{The time integrals of the nuclear/QSO luminosity function}
\label{sec:T}
\noindent
By integrating equation (\ref{eq:Euler}) over $L$ from 0 to $\infty$,
we have the conservation of the number density
\be
N(t_i,M\bhpr,t)\equiv\int_0^\infty{\cal N}(t_i,M\bhpr,L,t)\d L
\label{eq:Ndef}
\ee
that is,
\be
\frac{\partial N(t_i,M\bhpr,t)}{\partial t}=0.
\label{eq:Lagrangian}
\ee
According to equation (\ref{eq:Lagrangian}), we have
\be
N(t_i,M\bhpr,t)=N(t_i,M\bhpr,t_0).
\label{eq:consv},
\ee
For the same BH mass $M\bhpr$ at present,
we assume that their nuclear luminosity evolution is a function only of
the age of their nuclear activities $\tau\equiv t-t_i$.
Thus, using equations (\ref{eq:Ndef}) and (\ref{eq:consv}), we have
\begin{eqnarray}
{\cal N}(t_i,M\bhpr,L,t)
&=& N(t_i,M\bhpr,t)\delta [L-{\cal L}(M\bhpr,\tau)] \nonumber \\
&=& N(t_i,M\bhpr,t_0)\delta [L-{\cal L}(M\bhpr,\tau)],
\label{eq:calN}
\end{eqnarray}
where $\delta(x)$ is the Dirac function
and ${\cal L}(M\bhpr,\tau)$ is the nuclear luminosity of the progenitor of
the BH $M\bhpr$ at age $\tau$. 
As seen from equations (\ref{eq:nL}) and (\ref{eq:calN}), the nuclear
luminosity evolution (incorporated in the $\delta$-function) and the nuclear
activity triggering history [$N(t_i,M\bhpr,t_0)$] jointly contribute to the
nuclear LF.
By first integrating equation (\ref{eq:calN}) over $t_i$ from 0 to $t$ and
over $t$ from 0 to $t_0$ and then changing
the integration variables $(t,t_i)$ to $(t_i,\tau)$, we have
\begin{eqnarray}
\int_0^{t_0}\d t \int_0^t\d t_i~{\cal N}(t_i,M\bhpr,L,t)
&=& \int_0^{t_0} \d t \int_0^t \d t_i
~N(t_i,M\bhpr,t_0)\delta [L-{\cal L}(M\bhpr,\tau)] \nonumber \\
&=& \int_0^{t_0} \d t_i~N(t_i,M\bhpr,t_0)
\int_0^{t_0-t_i} \d \tau~\delta [L-{\cal L}(M\bhpr,\tau)] 
\nonumber \\
&=& \int_0^{t_0} \d t_i~N(t_i,M\bhpr,t_0)
\sum_k \frac{1} {\left|\d {\cal L}(M\bhpr,\tau)/\d \tau|_{\tau=\tau_k}\right|},
\label{eq:deriv}
\end{eqnarray}
where $\tau_k(L,M\bhpr)$ ($k=1,2,...$) are the roots of the equation
${\cal L}(M\bhpr,\tau)-L=0$ ($0<\tau<t_0-t_i$).\footnote{In equation
(\ref{eq:deriv}), it is assumed that
$\d {\cal L}(M\bhpr,\tau)/\d \tau \neq 0$ at $\tau=\tau_k$ ($k=1,2,..$).
Formulae (\ref{eq:deriv}), (\ref{eq:tool1}), (\ref{eq:taulife}) and
(\ref{eq:PLM}) are not difficult to generalize even if
$\d {\cal L}(M\bhpr,\tau)/\d \tau=0$ at $\tau=\tau_k$,
and other formulae will not be changed.}
We assume that the nuclear activities of all the local galaxies are quenched
at present.
Thus, for local BHs with mass $M\bhpr$, they have the same roots
$\tau_k(L,M\bhpr)$ ($k=1,2,...$) for the same $L$ since they have experienced
the same evolution of ${\cal L}(M\bhpr,\tau)$ before their quenching,
even though their accretion onto their seed BHs may be triggered at different time $t_i$.
Hence, the sum term $\sum_k$ in equation (\ref{eq:deriv}) does not
depend on $t_i$, and we have
\begin{eqnarray}
\int_0^{t_0}\d t \int_0^t\d t_i~{\cal N}(t_i,M\bhpr,L,t)
=n_{M\bh}(M\bhpr,t_0) 
\sum_k\frac{1}{\left|\d {\cal L}(M\bhpr,\tau)/\d \tau|_{\tau=\tau_k}\right|},
\label{eq:tool1}
\end{eqnarray}
where 
\be
n_{M\bh}(M\bhpr,t_0)\equiv\int_0^{t_0}N(t_i,M\bhpr,t_0)\d t_i
\ee
is the local BHMF and $n_{M\bh}(M\bhpr,t_0)\d M\bhpr$ gives the number
density of local BHs with mass in the range
$M\bhpr\rightarrow M\bhpr+\d M\bhpr$.
The lifetime of the nuclear activity for the BH with current mass $M\bhpr$
can be expressed by
\begin{eqnarray}
\tau\life(M\bhpr)
& = &\int \d \tau \label{eq:inttau} \\
& = &\int \d L \sum_k
\frac{1}{\left|\d {\cal L}(M\bhpr,\tau)/\d \tau|_{\tau=\tau_k}\right|}.
\label{eq:taulife}
\end{eqnarray}
Note that the integration in equation (\ref{eq:inttau}) is over the period
when the nucleus is active, and equation (\ref{eq:taulife}) is also restricted
to the luminosity ${\cal L}(M\bhpr,\tau)$ that is taken as active (e.g., higher than a certain luminosity limit).
The definition of ``active'' of galactic nuclei may be different in different
contexts, and equations (\ref{eq:inttau}) and (\ref{eq:taulife}) (such as
their integration limits) may be adjusted to appropriate forms according to
different definitions.
During the period of the nuclear activity, the fraction of the time (or the
probability) with luminosity in the range $L\rightarrow L+\d L$ can be given by
\be
P(L|M\bhpr)\d L\equiv \frac{\d L}{\tau\life(M\bhpr)}
\sum_k\frac{1}
{\left|\d {\cal L}(M\bhpr,\tau)/\d \tau|_{\tau=\tau_k}\right|}
\label{eq:PLM}
\ee
with
\be
\int P(L|M\bhpr)\d L=1.
\label{eq:normPLM}
\ee
Applying equations (\ref{eq:taulife}) and (\ref{eq:PLM}) in equation
(\ref{eq:tool1}), we have
\be
\int_0^{t_0}\d t \int_0^t\d t_i~{\cal N}(t_i,M\bhpr,L,t)=
n_{M\bh}(M\bhpr,t_0)\tau\life(M\bhpr)P(L|M\bhpr).
\label{eq:inttticalN}
\ee
By integrating equation (\ref{eq:inttticalN}) over $M\bhpr$ and using equation
(\ref{eq:nL}), we have the time integral of the nuclear LF as follows:
\be
\int_0^{t_0} n_L(L,t)\d t=\int_0^\infty \d M\bhpr
~n_{M\bh}(M\bhpr,t_0)\tau\life(M\bhpr)P(L|M\bhpr)\equiv \cT\Llocal(L,t_0).
\label{eq:TLlocal}
\ee

We define the QSOLF $\Psi_L(L,t)$ so that $\Psi_L(L,t)\d L$
is the comoving number density of QSOs with luminosity in the range
$L\rightarrow L+\d L$ at cosmic time $t$, and we define the time integral of the
QSOLF as follows:
\be
\cT\LQSO(L,t_0)\equiv \int_0^{t_0} \Psi_L(L,t)\d t.
\label{eq:TLQSO}
\ee
The QSOLF in equation (\ref{eq:TLQSO}) includes the contribution from both
optically bright QSOs and any other QSOs that are obscured or not seen in
optical bands but might be detectable in other bands (see \citealt{Fab03}).
In this paper $\Psi_L(L,t)$ denotes the LF of the ``live'' QSOs in the
distant universe and $n_L(L,t)$ denotes the nuclear LF of the progenitors
of the local BHs.
Based on the cosmological principle, similar to So{\l}tan's argument (1982),
the QSOLF $\Psi_L(L,t)$ represents the evolution of the nuclear luminosity of
local BHs, i.e., 
\be
n_L(L,t)=\Psi_L(L,t).
\label{eq:cospri}
\ee
Thus, with equations (\ref{eq:TLlocal})--(\ref{eq:cospri}), we have
\be
\cT\LQSO(L,t_0)=\cT\Llocal(L,t_0).
\label{eq:Trelation}
\ee
The physical meaning of equations (\ref{eq:TLlocal}) and (\ref{eq:Trelation})
can be understood clearly as follows.
For the progenitor of each local BH with mass $M\bhpr$, the average
time that it has spent in the nuclear luminosity range $L\rightarrow L+\d L$
is $\tau\life(M\bhpr)P(L|M\bhpr)$,
and taking QSOs as the progenitors of the local BHs, the total time spent in the range $L\rightarrow L+\d L$ by
the progenitors of all the local BHs with mass $M\bhpr$ in a unit
comoving volume should just be the time integral of $\Psi_L(L,t)\d L$.
We note that the quantity of the time integral of the QSOLF has been
(implicitly) used before in the literature, such as in some BH mass density
relations between local BHs and QSOs or the definition of the QSO mean
lifetime, which will be further discussed in \S~\ref{sec:degeneracy},
and we also note that \citet{Blandford03} points out that a simple
model of accretion implies a quantitative relationship between the time
integral of the QSOLF and the local BHMF.
Relation (\ref{eq:Trelation}) established above on the time integral of the
QSOLF will be the base to constrain the luminosity evolution of QSOs and BH
growth in this paper.

In addition, note that in equation (\ref{eq:Trelation}) or (\ref{eq:TLlocal})
the luminosity $L$ is assumed to be only a function of $M\bhpr$ and $\tau$.
For the more complicated case that $L$ also depends on some other parameters,
we may take the lifetime and probability function in equation
(\ref{eq:TLlocal}) as the averaged result over other parameters or
generalize these equations by including other parameters.

\subsection{Mean lifetime, and total/partial BH mass densities}
\label{sec:degeneracy}
\noindent
Below we show with appropriate assumptions, how the definition of the mean
QSO lifetime and some relations on the BH mass density obtained in the
literature (e.g.,\ \citealt{S82,YT02,HCO03}) can be incorporated in the
framework established above:
\begin{itemize}
\item Mean QSO lifetime: the lifetime $\tau\life(M\bhpr)$ in equation
(\ref{eq:TLlocal}) also represents the lifetime of ``live'' QSOs whose central
BH masses will be $M\bhpr$ at present.
By integrating equation (\ref{eq:Trelation}) over $L$ and using equation
(\ref{eq:normPLM}), we have
\be
\int\d L\int_0^{t_0} \Psi_L(L,t)\d t=\int_0^\infty \d M\bhpr
~n_{M\bh}(M\bhpr,t_0)\tau\life(M\bhpr).
\label{eq:meantau0}
\ee
We may define the mean lifetime of QSOs as follows:
\be
\langle\tau\life\rangle\equiv
\frac{\int_0^\infty \d M\bhpr n_{M\bh}(M\bhpr,t_0)\tau\life(M\bhpr)}
{\int_0^\infty \d M\bhpr n_{M\bh}(M\bhpr,t_0) H[\tau\life(M\bhpr)]}
=\frac{\int\d L\int_0^{t_0} \Psi_L(L,t)\d t}
{\int_0^\infty \d M\bhpr n_{M\bh}(M\bhpr,t_0)H[\tau\life(M\bhpr)]},
\label{eq:meantau}
\ee
where equation (\ref{eq:meantau0}) is used and $H(x)$ is the Heaviside step
function defined by $H(x)=1$ if $x>0$ and $H(x)=0$ otherwise.
We denote the mass of the progenitors of the local BHs when they had a nuclear
luminosity $L$ (or the central BH mass of a QSO with luminosity $L$) as $M\bh$.
If $L$ is an increasing function only of $M\bh$ and only the period with
luminosities brighter than a certain value $L(M\bh')$ is taken as active phase,
we have $\tau\life(M\bhpr)>0$ if $M\bhpr>M\bh'$ and $\tau\life(M\bhpr)=0$
otherwise.
Thus, the mean lifetime of QSOs is given by
\be
\langle\tau\life\rangle
=\frac{\int_{L(M\bh')}^\infty\d L\int_0^{t_0} \Psi_L(L,t)~\d t}
{\int_{M\bh'}^\infty n_{M\bh}(M\bhpr,t_0)~\d M\bhpr},
\ee
which is identical to equation (59) in Yu \& Tremaine (2002;
see also eq.~[24] in \citealt{HCO03}).

\item Total BH mass densities:
we denote the QSO bolometric luminosity produced by a mass accretion rate
$\dot M\acc$ as $L\bol=\epsilon \dot M\acc c^2
=\epsilon \dot M\bh/(1-\epsilon)$, where
$\dot M\bh=\dot M\acc (1-\epsilon)$ is the growth
rate of BH mass and $c$ is the speed of light.
(The subscript ``bol'' represents the bolometric luminosity,
and the symbol $L$ in this paper may be either the bolometric
luminosity or the luminosity in a specific band, if not specified.)
By multiplying equation (\ref{eq:Trelation}) by
$(1-\epsilon)L\bol/(\epsilon c^2)=\dot M\bh$
and then integrating over $L$ from a given value $L'$ to $\infty$, we have
\be
\int_{L'}^\infty\d L \frac{(1-\epsilon)L\bol}{\epsilon c^2}
\int_0^{t_0} \Psi_L(L,t)\d t=\int_0^\infty \d M\bhpr
~n_{M\bh}(M\bhpr,t_0) \int_{L'}^\infty\d L~\dot M\bh
\tau\life(M\bhpr)P(L|M\bhpr).
\label{eq:soltan0}
\ee
By setting $L'=0$, equation (\ref{eq:soltan0}) becomes
\be
\int_0^\infty\d L \frac{(1-\epsilon)L\bol}{\epsilon c^2}
\int_0^{t_0} \Psi_L(L,t)\d t=\int_0^\infty \d M\bhpr~M\bhpr n_{M\bh}(M\bhpr,t_0),
\label{eq:soltan}
\ee
where
\be
\int_0^\infty\d L~\dot M\bh \tau\life(M\bhpr)P(L|M\bhpr)=
\int_0^{t_0}\dot M\bh~\d t=M\bhpr-\langle M\bhti\rangle(M\bhpr)
\label{eq:soltan1}
\ee
is used; $\langle M\bhti\rangle(M\bhpr)$ is the average mass of the seed BHs
that will have mass $M\bhpr$ at present and is ignored in equation
(\ref{eq:soltan}).
Equation (\ref{eq:soltan}) is identical to So{\l}tan's argument (1982)
relating the total local energy density in QSOs to the total BH mass density
in nearby galaxies.

\item Partial BH mass densities:
if the QSO luminosity $L$ is only an increasing function of the BH mass
$M\bh$, we have
\begin{eqnarray}
& & \int_{L'}^\infty\d L~\dot M\bh \tau\life(M\bhpr)P(L|M\bhpr)\nonumber \\
&=& \cases {M\bhpr-\max[M\bh(L'),M\bhti(M\bhpr)],
                          & if $M\bhpr>\max[M\bh(L'),M\bhti(M\bhpr)] $ \cr
        0                , & otherwise},
\label{eq:partial1}
\end{eqnarray}
where the seed BH mass $M\bhti$ is assumed to be a function only of $M\bhpr$.
By applying equation (\ref{eq:partial1}) in equation (\ref{eq:soltan0})
and assuming $M\bhti(M\bhpr)<M\bh(L')$, we have
\be
\int_{L'}^\infty\d L \frac{(1-\epsilon)L\bol}{\epsilon c^2}
\int_0^{t_0} \Psi_L(L,t)\d t=\int_{M\bh(L')}^\infty
\d M\bhpr~[M\bhpr-M\bh(L')]n_{M\bh}(M\bhpr,t_0).
\label{eq:YT}
\ee
Actually, even if the seed BH mass is not a function only of $M\bhpr$,
equation (\ref{eq:YT}) will still hold at a given $L'$ as long as all the seed
BH masses of the local BHs $M\bhpr$[$>M\bh(L')$] are smaller than $M\bh(L')$.
Equation (\ref{eq:YT}) is identical to the relation between
the partial BH mass density in nearby galaxies and that accreted during
bright QSO phases in \citet{YT02} if BH mergers are ignored
(see eq.~30 in \citealt{YT02}).
\end{itemize}

\subsection{${\cal L}(M\bhpr,\tau)$} \label{sec:Ltauage}

\noindent
Given the QSOLF and local BHMF, the nuclear luminosity evolution
[or $\tau\life(M\bhpr)P(L|M\bhpr)$] cannot be uniquely determined
from equations (\ref{eq:TLlocal})--(\ref{eq:Trelation}).
Below we use some physical arguments to assume a form of the
luminosity evolution ${\cal L}(M\bhpr,\tau)$.
The parameters involved in the assumed form will be constrained
by observations in \S~\ref{sec:constraint}.

It is generally believed that the growth of a BH involves two phases after
the nuclear activity is triggered on
(see the discussion on the ``feast and famine'' model in \citealt{SB92} and
\citealt{Blandford03}).
In the first (or ``demand limited'') phase, there is plenty of material to
supply for the BH growth;
however, not all of the available material can contribute to the BH growth
at once and the BH growth is limited by the Eddington luminosity.
With the decline of material supply, the BH growth enters into the second
(or ``supply limited'') phase and the nuclear luminosity is expected to
decline below the Eddington luminosity.
For simplicity, we assume that the two phases appear only once in this
paper, although this assumption is not required in relation (\ref{eq:Trelation}).
The possibility of more complicated accretion patterns
deserves further investigation by applying corresponding models in
relation (\ref{eq:Trelation}).

In the first phase, after the BH is triggered at time $t_i$, we assume that
it accretes with the Eddington luminosity for a time $\tau\I$
and its mass increases to be $M\bhtI$ at time $t=t_i+\tau\I\equiv t\I$.
The mass-to-energy conversion efficiency $\epsilon$ is assumed to be a
constant. Thus, the nuclear luminosity in the first phase increases with
time as follows:
\be
{\cal L}\bol(\tau)=L\Edd(M\bhtI)
\exp\left(\frac{\tau-\tau\I}{\tau\Sp}\right),
\qquad 0<\tau<\tau\I,
\label{eq:Lphase1}
\ee
where $L\Edd(M\bh)$ is the Eddington luminosity of a BH with mass $M\bh$ and
\be
\tau\Sp=4.5\times 10^7 \left[\frac{\epsilon}{0.1(1-\epsilon)}\right]\yr
\label{eq:tauSp}
\ee
is the Salpeter time (the time for a BH radiating at the Eddington luminosity
to e-fold in mass).
In the second phase, we assume that the evolution of the nuclear luminosity
declines as follows (e.g.,\ \citealt{HNR98,HL98}):
\be
{\cal L}\bol(\tau)=\cases{L\Edd(M\bhtI)\exp\left(-\frac{\tau-\tau\I}{\tau\D}\right), & for $\tau\I\le\tau\le\tau\I+\xi\tau\D$, \cr
0, & for $\tau>\tau\I+\xi\tau\D$,}
\label{eq:Lphase2}
\ee
where $\tau\D$ is the characteristic declining timescale of the nuclear
luminosity.
We assume that QSOs become quiescent when the nuclear luminosity declines by
a factor of $\eta=\exp(-\xi)$ compared to the peak luminosity $L\Edd(M\bhtI)$,
so there is a cutoff of the nuclear luminosity at
$\tau=\tau\I+\xi\tau\D$ in equation (\ref{eq:Lphase2}).
The factor $\xi$ is set to $-\ln(10^{-3})=6.9$ here, since after
decreasing by a factor of $\eta=10^{-3}$, the nuclear luminosity of
BHs even with a high mass $\sim 10^9\msun$ will become fainter than the
luminosity range ($M_B\la -20$ in \S~\ref{sec:comparison}) of
interest in this paper.
With the assumption that all QSOs are quenched at present
(i.e., $t_0-t_i-\tau\I \gg \tau\D$), the BH mass at present is given by
\be
M\bhpr=M\bhtI+\int_{t\I}^{t_0}\frac{(1-\epsilon)L\bol}{\epsilon c^2}\d t
=\left(1+\frac{\tau\D}{\tau\Sp}\right)M\bhtI.
\label{eq:Mbhpr}
\ee
In equation (\ref{eq:Mbhpr}), the efficiency $\epsilon$ is assumed to be the
same as the efficiency in the
first phase and not to change with the decline of the nuclear luminosity.
It is important to know how the realistic efficiency evolves with the change
of the nuclear luminosity or other parameters (which would depend on the
evolution of both the accretion rates and BH spins; a detailed study about this
is beyond the scope of this paper).  If the nuclear luminosity of a BH is
smaller than its Eddington luminosity by a factor of 10 or more, according to
current accretion models, the BH might accrete material via the advection-dominated
accretion flow (ADAF) with low efficiency $\epsilon \ll 0.1$ (e.g.,\
\citealt{NY94}), rather than via the thin-disk accretion with efficiency
$\epsilon\sim 0.1-0.3$ near the Eddington luminosity. Considering this
possibility of the very low efficiency at the late stage of the luminosity
evolution, the contribution from the very low efficiency stage to the time
integral of the nuclear LF (in the luminosity range $M_B<-20$ shown in this
paper) is insignificant and our conclusions will still hold for other stages of
the nuclear activity (which can be inferred from the results obtained in
\S~\ref{sec:constraint}).  We also assume that $\epsilon$ is irrelevant to the
BH mass $M\bhpr$ in this paper.

By using equations (\ref{eq:Lphase1})--(\ref{eq:Mbhpr}) in equations
(\ref{eq:taulife}) and (\ref{eq:PLM}), we have the QSO lifetime given by
\be
\tau\life=\tau\I+\xi\tau\D,
\label{eq:tauQSO}
\ee
and the BH with present mass $M\bhpr$ has such a probability distribution of
the nuclear bolometric luminosity in its evolution history:
\be
P(L\bol|M\bhpr)=\frac{1}{\tau\life}\left(f\I\frac{\tau\Sp}{L\bol}+f\D\frac{\tau\D}{L\bol}\right),
\qquad
\label{eq:PLMQSO}
\ee
where
\be
f\I=\cases {1, & if $L\Edd(M\bhtI)\exp(-\tau\I/\tau\Sp)\le L\bol \le
L\Edd(M\bhtI)$
\cr 0, & otherwise},
\label{eq:fI}
\ee
and
\be
f\D=\cases {1, & if $L\Edd(M\bhtI)\exp(-\xi)\le L\bol \le L\Edd(M\bhtI)$ \cr
0, & otherwise}.
\label{eq:fD}
\ee
In the analysis above, the timescales $\tau\I$ and $\tau\D$ are not
necessarily constants, and they may be a function of $M\bhpr$.
The dependence of $\tau\I$ on $M\bhpr$ would depend on the distribution of
seed BHs, which are poorly known and beyond the scope of this paper.
Below we always assume that $\tau\I$ and $\tau\D$ are irrelevant to
$M\bhpr$.

\section{The local BH mass function and QSO luminosity function obtained from
observations}\label{sec:obs}

\subsection{The local BH mass function $n_{M\bh}(M\bhpr,t_0)$}
\label{sec:BHMF}
\noindent
In this subsection we use the velocity-dispersion distribution of
local galaxies (\S~\ref{sec:nsigma}) and the empirical BH mass-velocity
dispersion relation (\S~\ref{sec:msigma}) to obtain the local BHMF
(\S~\ref{sec:BHMFres}).

\subsubsection{Local velocity-dispersion function $n_\sigma(\sigma,t_0)$}
\label{sec:nsigma}
\noindent
We define $n_\sigma(\sigma,t_0)$ as the velocity-dispersion function of
the hot stellar components of local galaxies so that
$n_\sigma(\sigma,t_0)\d\sigma$ represents the comoving number density of
local galaxies in the range $\sigma\rightarrow\sigma+\d\sigma$
(by ``hot''component we mean either an elliptical galaxy or the bulge of
a spiral or S0 galaxy).
The velocity-dispersion distribution $n_\sigma(\sigma,t_0)$ includes
the contribution by both early-type galaxies $n\early_\sigma(\sigma,t_0)$
and late-type galaxies $n\late_\sigma(\sigma,t_0)$, that is,
\be
n_\sigma(\sigma,t_0)=n\early_\sigma(\sigma,t_0)+n\late_\sigma(\sigma,t_0).
\label{eq:veldisp}
\ee

Recent study on a sample of nearly 9000 nearby early-type galaxies obtained
by the SDSS has provided the velocity-dispersion distribution in early-type
galaxies as follows
(see eq.~4 in \citealt{Sheth03}, and \citealt{Bernardi03}):
\noindent
\be
n\early_\sigma(\sigma,t_0)=\phi_*\left(\frac{\sigma}{\sigma_*}\right)^{\alpha}
\frac{\exp\left[-(\sigma/\sigma_*)^{\beta}\right]}{\Gamma(\alpha/\beta)}
\frac{\beta}{\sigma},
\label{eq:earlydisp}
\ee
where the best-fit values of $(\phi_*,\sigma_*,\alpha,\beta)$ are 
$(0.0016\pm0.0001,88.8\pm17.7,6.5\pm1.0,1.93\pm0.22)$,
$\phi_*$ is the comoving number density of local early-type galaxies in units
of $\Mpc^{-3}$, and $\sigma_*$ is in units of $\kms$.

We obtain the velocity-dispersion distribution in local late-type galaxies
$n_\sigma\late(\sigma,t_0)$ in the following steps:
(i) Following \citet{Sheth03}, we get the LF of the late-type
galaxies by subtracting the LF of the early-type galaxies
\citep{Blanton03} from the LF of total galaxies
\citep{Bernardi03}.
(ii) Following \citet{Sheth03}, we get the distribution of the circular
velocity $v\c$ in late-type galaxies by using the LF of the
late-type galaxies obtained above and the following Tully-Fisher relation
\citep{Giovanelli}:
\be
\log(2v\c)=1.10-(M_I-5\log h)/7.95,
\label{eq:tullyfisher}
\ee
where $M_I$ is the absolute magnitude of the galaxies in the $I$ band,
accounting for the intrinsic scatters around relation
(\ref{eq:tullyfisher}) and the inclination effects of galaxies
(see details in \citealt{Sheth03}).
(iii) We get the velocity-dispersion function of late-type galaxies
by using the circular-velocity distribution of the late-type galaxies
obtained above and the following correlation between the circular velocity
and the velocity dispersion of the bulge component
(see eq.~3 in \citealt{Baes03}; see also \citealt{Ferrarese02}):
\be
\log\left(\frac{v\c}{200\kms}\right)=
(0.96\pm0.11)\log\left(\frac{\sigma}{200\kms}\right)+(0.21\pm0.023).
\label{eq:vcircsigma}
\ee
The intrinsic scatter of the correlation given by (\ref{eq:vcircsigma}) is small 
($<0.15\dex$; see Fig.~1 in \citealt{Baes03}) and ignored in our
calculation. 
Relation (\ref{eq:vcircsigma}) may not hold for $\sigma<80\kms$, which
corresponds to BH mass $\la 4\times 10^6\msun$ according to the
BH mass-velocity dispersion relation below (eq.~\ref{eq:msigma}) and
is beyond the range that we focus on in \S~\ref{sec:constraint}.

\subsubsection{The BH mass and velocity dispersion relation}
\label{sec:msigma}
\noindent
Studies of central BHs in nearby galaxies have revealed that the BH mass
and the velocity dispersion of the hot stellar component of the
host galaxy follow a tight correlation \citep{Tremaine02,FM00,Gebhardt00}.
The logarithmic of the BH mass at a given velocity dispersion $\sigma$
has a mean value given by \citep{Tremaine02}
\be
\langle \log (M\bhpr)|\sigma\e \rangle
=A+\gamma\log(\sigma\e/200\kms),
\label{eq:msigma}
\ee
where $M\bhpr$ is in units of $\msun$, $\gamma=4.02\pm0.32$,
$A=8.18\pm0.06$ have been adjusted to
our assumed Hubble constant $h=0.65$ (see section 2.2 in \citealt{YT02}),
and $\sigma\e$ is the luminosity-weighted line-of-sight velocity dispersion
within a slit extending to the effective radius.
Note that $\sigma\e$ in the correlation (\ref{eq:msigma}) is the velocity
dispersion within a slit extending to the effective radius $R_o$
\citep{Gebhardt00},
while $\sigma$ in the SDSS (eq.~\ref{eq:earlydisp}) is the velocity dispersion
within a circular aperture extending to $R_o/8$.
However, replacing $\sigma\e$ with $\sigma$ in equation (\ref{eq:msigma})
will not cause much difference as the two definitions
should give very similar results \citep{Tremaine02}.
\citet{MF01a} give an alternative version of the correlation 
(\ref{eq:msigma}) with a steeper slope, $\gamma=4.72$.
The reasons for this difference in slopes are discussed by \citet{Tremaine02}.
We perform the calculations using both versions of the
correlation and the difference on the local BHMF will be
discussed in \S~\ref{sec:BHMFres}.

Note that relation (\ref{eq:msigma}) is fitted in the
$\log M\bhpr$--$\log\sigma$ space.
We assume that the distribution in $\log M\bhpr$ at a given $\sigma$ is
Gaussian with intrinsic standard deviation
$\scatter$, which is independent of $\sigma$ and thus can be written as
\be
P(\log M\bhpr|\sigma)=\frac{1}{\sqrt{2\pi}\scatter}
\exp\left[-\frac{(\log M\bhpr-\langle \log M\bhpr|\sigma\rangle)^2}
{2\Delta^2_{\log M\bhpr}}\right].
\label{eq:probm0tom}
\ee
According to \citet{Tremaine02}, the intrinsic scatter in $\log M\bhpr$ 
should not be larger than 0.25--$0.3\dex$.
Nevertheless, to check the effect of $\scatter$ on the results, below we
show the results obtained with a value of $\scatter$($=0.4\dex$) higher than
$0.27\dex$, as well as those obtained with $\scatter=0$ and $0.27\dex$.

With the velocity-dispersion function and the BH mass--velocity dispersion
relation (eqs.~\ref{eq:msigma} and \ref{eq:probm0tom}), the local BHMF
is given by
\be
n_{M\bh}(M\bhpr,t_0)=\ln 10
\int M\bhpr P(\log M\bhpr|\sigma)n_\sigma(\sigma,t_0)
\d\sigma.
\label{eq:bhmassfunc}
\ee

\subsubsection{Results: $n_{M\bh}(M\bhpr,t_0)$} \label{sec:BHMFres}
\begin{figure}
\epsscale{0.55}
\plotone{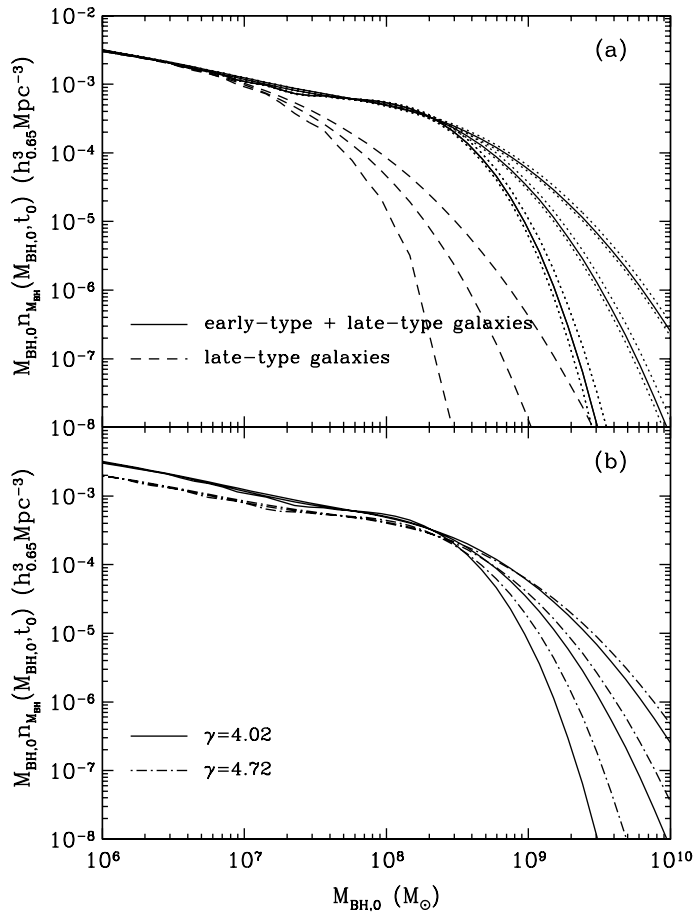}
\caption{Local BHMF obtained from observations of the velocity
dispersion distribution of nearby galaxies and
the BH mass-velocity dispersion relation.
(a) Solid lines represent the $M\bhpr n_{M\bh}(M\bhpr,t_0)$ in both
early-type and late-type galaxies
(see eq.~\ref{eq:veldisp}--\ref{eq:bhmassfunc});
dashed lines represent the $M\bhpr n_{M\bh}(M\bhpr,t_0)$ in late-type
galaxies.
Different solid or dashed lines are obtained
with different values of the intrinsic scatter $\scatter$
(=0, 0.27, and $0.4\dex$ from bottom to top at the high-mass end, respectively)
in the $M\bhpr-\sigma$ relation (eq.~\ref{eq:msigma}).
The two dotted lines adjacent to each solid line give the effect of the
1-$\sigma$ error of the best-fit parameters in the velocity-dispersion function
of early-type galaxies (eq.~\ref{eq:earlydisp}).
(a) shows that the local BHMF at the high-mass end
($\ga 3\times 10^8\msun$) is significantly affected by $\scatter$,
and the uncertainty of the BHMF due to the
uncertainty in the velocity-dispersion function of early-type galaxies
is negligible compared to the uncertainty due to the intrinsic scatter
$\scatter$ in the $M\bhpr-\sigma$ relation.
(b) Solid lines are the same as those in (a), which are
obtained with the slope $\gamma=4.02$ in relation (\ref{eq:msigma})
\citep{Tremaine02}, while the dot-dashed lines are obtained with
$\gamma=4.72$ \citep{MF01a}.
The difference of the BHMF at the high-mass end due to the difference
in $\gamma$ is not as significant as that due to the difference of the
intrinsic scatter $\scatter$ in the $M\bhpr-\sigma$ relation.
See details in \S~\ref{sec:BHMFres}.
}
\label{fig:lmf}
\end{figure}

\noindent
Using the velocity-dispersion function in \S~\ref{sec:nsigma},
the BH mass--velocity dispersion relation in \S~\ref{sec:msigma}, and equation
(\ref{eq:bhmassfunc}), we obtain the local BHMF $n_{M\bh}(M\bhpr,t_0)$
and show the distribution of $M\bhpr n_{M\bh}(M\bh,t_0)$ in
Figure~\ref{fig:lmf}.
In Figure~\ref{fig:lmf}a, different solid lines represent the
results obtained with different $\scatter$ (=0, 0.27, and $0.4\dex$ from bottom
to top at the high-mass end, respectively, see eq.~\ref{eq:probm0tom}).
As seen from panel (a), the local BHMF at the high-mass end
($\ga 3\times 10^8\msun$) is significantly affected by the
intrinsic scatter $\scatter$.
For example, the BHMF obtained with $\scatter=0.27\dex$
(middle solid line) is larger than that obtained with $\scatter=0$
(bottom solid line) by a factor of 5 at $M\bhpr\simeq 10^9\msun$ and
by a factor of more than 10 at $M\bhpr\simeq 4\times 10^9\msun$
and if $\scatter=0.4\dex$,
the BHMF (top solid line) is larger than that obtained
with $\scatter=0$ by a factor of 8 at $M\bhpr\simeq 10^9\msun$
and a factor of more than 100 at $M\bhpr\simeq 4\times 10^9\msun$.
The solid lines are the results obtained by setting the variables
$(\phi_*,\sigma_*,\alpha,\beta)$ in equation (\ref{eq:earlydisp})
to the mean of the best-fit values.
We also show the uncertainty of the local BHMF due to the 1-$\sigma$ error
of the
fitting values in the velocity-dispersion function of early-type galaxies
in panel (a).
The two dotted lines adjacent to each solid line represent the result after
considering the 1-$\sigma$ error of the mean value of
$(\sigma_*,\alpha,\beta)$ in equation (\ref{eq:earlydisp}),
which is obtained by setting
$(\sigma_*,\alpha,\beta)=(88.8,6.5,1.93)\pm(+17.7,-1.0,+0.22)$.
Note that the signs set to the errors of the variables are
not all the same because the best-fit values of $(\sigma_*,\alpha,\beta)$
are strongly correlated with one another (see fig.~4 in \citealt{Sheth03}).
We do not consider the error in $\phi_*$ here since it is small and
the uncertainty caused by it is negligible compared to that caused by the
error of other variables.
As seen from panel (a), the uncertainty of the local BHMF due to the
uncertainty in the velocity-dispersion function of early-type galaxies
is negligible compared to that due to the intrinsic scatter $\scatter$
in the $M\bhpr-\sigma$ relation.
In panel (a), the dashed lines show the local BHMF in late-type galaxies
also obtained with $\scatter=$0, 0.27 and $0.4\dex$ from bottom to top at
the high-mass end, respectively.
So far, it is hard to give an accurate estimate of the uncertainty in the
estimate of the velocity-dispersion function of late-type galaxies here.
We believe that the effect on the local BHMF due to the
uncertainty in the velocity-dispersion function of the late-type galaxies
can be ignored at the high-mass end ($\ga 4\times 10^7\msun$),
where the local BHMF is dominated by early-type galaxies.
The local BHMF is dominated by late-type galaxies
at the low-mass end ($\la 10^7\msun$).
We expect that the local BHMF with BH mass in the range
from $4\times 10^6\msun$ to $4\times 10^7\msun$ has considerable
accuracy, for example, within 50\%.

In Figure~\ref{fig:lmf}b
we show the effect on the local BHMF due to different versions of the slope
$\gamma$ in the $M\bhpr-\sigma$ relation (eq.~\ref{eq:msigma}).
The solid lines are the same as those in panel (a), which are obtained
with $\gamma=4.02$ \citep{Tremaine02}.
The dot-dashed lines are obtained with $\gamma=4.72$ \citep{MF01a}.
As seen from panel (b), the difference of the BHMF at the high-mass
end due to the difference in $\gamma$ is not so significant as that due to
the difference of the intrinsic scatter $\scatter$ in the $M\bhpr-\sigma$
relation.
Below we always show the results obtained with $\gamma=4.02$ \citep{Tremaine02}.

In principal, the local BHMF can also be obtained by using the
luminosity distribution of the hot stellar components of local galaxies and
its correlation with central BHs (e.g.,\ \citealt{KG01}). However, we do not
do so in this paper, because the total local BH mass density obtained by using
the BH mass -- luminosity relation appears to be larger than that obtained by
using the BH mass -- velocity dispersion relation by a factor of more than 2,
and the reason for this difference is not yet clear \citep{YT02}.
As argued in \citet{YT02}, we believe that the results obtained by using the
BH mass -- velocity dispersion relation are more reliable, as the correlation
between the BH mass and the velocity dispersion is tighter and the result does
not depend on the uncertain bulge-disk decomposition. 

Note that although the local BHMF at the high-mass end is
significantly affected by the intrinsic scatter in the $M\bhpr-\sigma$
relation, the total local BH mass density is not,
which is higher than the total density obtained by setting $\scatter=0$
only by a factor of $\exp[{1\over2}(\scatter\ln10)^2]$(=1.2 if
$\scatter=0.27\dex$ and 1.5 if $\scatter=0.4\dex$
(see eq.~12 in \citealt{YT02}).

\subsection{The luminosity function of optically bright QSOs}
\label{sec:QSOLF}
\noindent 
The LF of optically bright QSOs is often fitted with a double power law:
\be
\Psi_{M_B}\opt(M_B,z)=\frac{\Psi_M^*}
{10^{0.4(\beta_1+1)[M_B-M^*_B(z)]}+10^{0.4(\beta_2+1)[M_B-M^*_B(z)]}},
\label{eq:QSOLF}
\ee
where the superscript ``opt'' represents optically bright (or unobscured)
QSOs, $\Psi_{M_B}\opt(M_B,z)dM_B$ is the comoving number density of QSOs with
absolute magnitude in the range [$M_B, M_B+dM_B$] at redshift $z$, and
we have 
\be
\Psi_{M_B}(M_B,z)=\Psi_{L_B}(L_B,z)|dL_B/dM_B|=0.92L_B\Psi(L_B,z).
\ee
\citet{Pei95} uses this form to fit the data set from \citet{HS90}
and \citet{WHO95} on the basis of more than 1200 QSOs over the range of
redshift $0.1<z<4.5$.
In the cosmological model of $(\Omega_{\rm m},\Omega_\Lambda,h)=(1,0,0.5)$,
the QSOLF with absolute magnitudes $-30\la M_B\la -23$ can
be fitted by equation (\ref{eq:QSOLF}) with the following parameters:
\begin{eqnarray}
\Psi_M^*=6.7\times10^{-6}h^3\Mpc^{-3}\mag^{-1},\label{eq:phistarpei}\\
M^*_B(z)=M^*_B(0)+1.25\log(1+z)-2.5(k_1z+k_2z^2), \label{eq:Mstarzpei}\\
M^*_B(0)=-20.83+5\log h, k_1=1.39, k_2=-0.25, \label{eq:k1k2pei}\\
\beta_1=-1.64 ~~{\rm and~~}  \beta_2=-3.52. \label{eq:betapei}
\label{eq:QSOLFparapei}
\end{eqnarray}
Note that the cosmological model in which the parameters
above are fitted is different from the model adopted in this paper.
In our calculations below, we have transferred the parameters in
equations (\ref{eq:phistarpei})--(\ref{eq:QSOLFparapei}) into the cosmological
model used in this paper, $(\Omega_{\rm m},\Omega_\Lambda,h)=(0.3,0.7,0.65)$.

\citet{Boyle00} use the function form of equation~(\ref{eq:QSOLF}) to fit
a much larger data set from the 2dF QSO redshift survey \citep{Boyle00}
and Large Bright QSO survey \citep{HFC95}, which contain over 6000 QSOs,
and give the QSOLF with absolute magnitudes $-26<M_B<-23$
and redshift $0.35<z<2.3$ in our standard cosmological model
$(\Omega_{\rm m},\Omega_\Lambda)=(0.3,0.7)$ by the following parameters:
\begin{eqnarray}
\Psi_M^*=2.9\times10^{-6}h^3\Mpc^{-3}\mag^{-1},\label{eq:phistar}\\
M^*_B(z)=M^*_B(0)-2.5(k_1z+k_2z^2), \label{eq:Mstarz}\\
M^*_B(0)=-21.14+5\log h, k_1=1.36, k_2=-0.27, \label{eq:k1k2}\\
\beta_1=-1.58 {~~\rm and~~}  \beta_2=-3.41. \label{eq:beta}
\label{eq:QSOLFpara}
\end{eqnarray}

The quadratic dependence of the characteristic magnitude $M^*_B(z)$ on $z$
in equations (\ref{eq:Mstarzpei}) and (\ref{eq:Mstarz})
shows an increasing characteristic luminosity with increasing redshift at
low redshift ($z\la 2.5$) and a decline of the characteristic luminosity
at higher redshift, which is suggested by observations
(e.g. Shaver et al. 1996), although the LF is not yet
accurate enough to confirm the decline at $z>2.5$.

The QSOLF over the range $3.6<z<6$ provided in
\citet{Fan01,Fan03} gives a flatter bright-end slope ($\beta_2=-2.5$)
than equations (\ref{eq:QSOLFparapei}) and (\ref{eq:QSOLFpara});
however, in our calculation below, we simply
extrapolate equations (\ref{eq:QSOLFparapei}) and (\ref{eq:QSOLFpara})
to high redshift because the detailed QSOLF at $z>3.5$
does not affect our results much (see Fig.~\ref{fig:philf} below;
if we use the LF in \citealt{Fan01}, our results change by
less than a few percent).

Using equations (\ref{eq:QSOLF})--(\ref{eq:QSOLFpara}), we obtain the time
integral of the LF of optically bright QSOs
\be
\cT\MBQSOopt\MBpr\equiv \int_0^\infty\Psi\MBQSOopt(M_B,t)\d t
=0.92L_B\cT\LBQSOopt(L_B,t_0)
\label{eq:TMBQSO}
\ee
as a function of the absolute magnitude $M_B$ shown in Figure~\ref{fig:philf}.
Both of the results obtained from the QSOLFs given by Boyle et al. (2000;
shown as solid lines; eqs.~\ref{eq:phistarpei}--\ref{eq:QSOLFparapei})
and those given by Pei (1995; shown as dotted lines;
eqs.~\ref{eq:phistar}--\ref{eq:QSOLFpara}) are presented.
In the calculation, the QSOLFs are extrapolated to the parameter space
not covered by the observations.
In Figure 2a we show the integral of the QSOLF over time $0\le z \le 2.3$ as
the bottom solid and dotted lines and the integral of the QSOLF over the whole
cosmic time $\cT\MBQSOopt\MBpr$ as the top solid and dotted lines.
As seen from panel (a), the difference of the time integral of the QSOLF due
to different versions of QSOLFs is negligible.
In  Figure 2b we show the fraction of $\cT\MBQSOopt\MBpr$ contributed by
QSOs in the range $0\le z\le2.3$ (bottom solid and dotted lines),
which is more than 50\% at $M_B>-26$ and 75\% at $M_B>-23$,
and we also show the fraction contributed by QSOs in the range
$0\le z\le3.5$ (top solid and dotted lines),
which is more than 95\% at any $M_B$ in the figure.
Thus, the effect of the uncertainty in the QSOLF at high redshift on the
time integral of QSOLF can be neglected.

\begin{figure}
\begin{center}
\epsscale{0.7}
\plotone{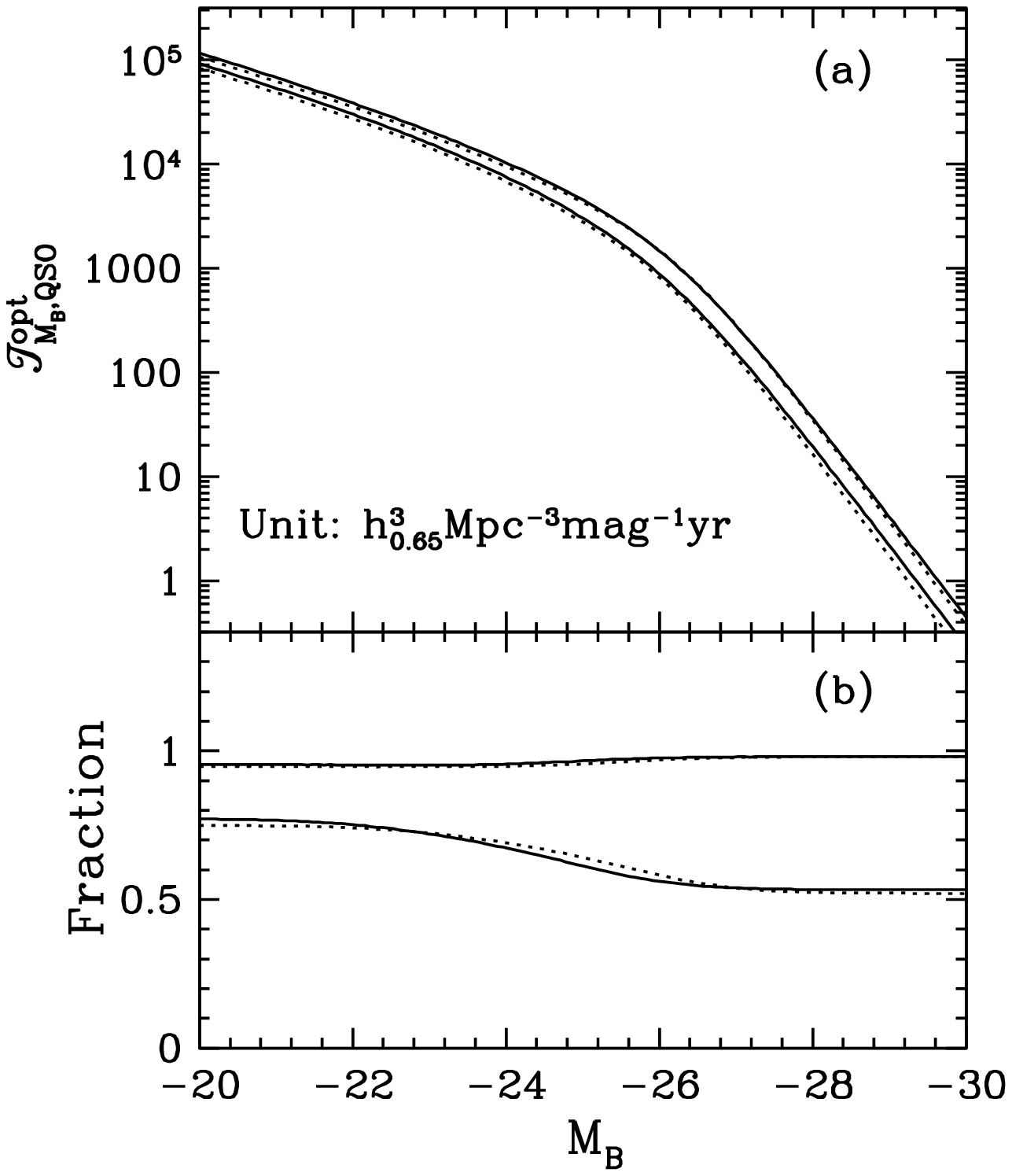}
\caption{Time integral of the LF of optically bright QSOs as a function of
the absolute magnitude $M_B$ (see eq.~\ref{eq:TMBQSO}). 
The results obtained by using the QSOLFs given by \citet{Pei95}
(eqs.~\ref{eq:phistarpei}--\ref{eq:QSOLFparapei}) are shown as dotted lines,
and those obtained by using the QSOLFs
given by \citet{Boyle00} (eqs.~\ref{eq:phistar}--\ref{eq:QSOLFpara}) 
are shown as solid lines.
The redshift range of the QSOLF is only $0.35<z<2.3$ for \citet{Boyle00}
and $0.1<z<4.5$ for \citet{Pei95}.
In the calculations, both are extrapolated to the parameter space 
not covered by the observations.
(a) Top solid and dotted lines represent the integral of the
QSOLF over the whole cosmic time $\cT\MBQSOopt\MBpr$;
bottom solid and dotted lines represent the integral of the QSOLF
over time in the redshift range $0\le z\le 2.3$.
The difference due to different versions of the QSOLFs is negligible.
(b) Top solid and dotted lines give the fraction of
$\cT\MBQSOopt\MBpr$ contributed by the QSOs with redshift
$0\le z\le 3.5$;
bottom solid and dotted lines give the fraction contributed by the
QSOs with redshift $0\le z \le 2.3$.
This figure shows that the effect due to the uncertainty of the QSOLF at high
redshift on the time integral of QSOLF can be neglected.
See \S~\ref{sec:QSOLF}.
}
\label{fig:philf}
\end{center}
\end{figure}

\section{Is the local BH mass function consistent with the QSO luminosity
function?}\label{sec:comparison}
\noindent
With the goal to obtain accurate observational constraints on the QSO
luminosity evolution and the fundamental parameters of the QSO model
(see \S~\ref{sec:Ltauage}), we compare the time-integral of the QSOLF
with that predicted from local BHs in this section (see eq.~\ref{eq:Trelation}
in \S~\ref{sec:T} and inequalities~\ref{eq:TMBrelation} and \ref{eq:Ropt}
below).
Note that the luminosity in the assumed model on the QSO luminosity
evolution in \S~\ref{sec:Ltauage} is the bolometric luminosity $L\bol$.
However, the QSOLF obtained from observations is usually
in a specific band (which is the $B$ band in this paper).
Before the comparison, we first transfer the
bolometric luminosity to the luminosity in a specific band.

\subsection{The bolometric correction}\label{sec:bol}
\noindent
We denote $C_B$ as the bolometric correction in the $B$ band,
defined through $L\bol\equiv C_B L_{\nu_B}$, where $L_{\nu_B}$ is the energy
radiated at the central frequency of the $B$ band per unit time and
logarithmic interval of frequency.
We assume that $C_B$ is independent of the cosmic time $t$.
We denote $P(C_B|L\bol)$ as the probability distribution function of the
bolometric correction $C_B$ at a given $L\bol$.
Thus, we have the LF of the progenitors of local BHs
in the $B$ band as follows:
\be
n_\LnuB(\LnuB,t)=\int C_B P(C_B|L\bol) n_{L\bol}(L\bol,t)~\d C_B.
\label{eq:plmft11}
\ee
With equations (\ref{eq:TLlocal}) and (\ref{eq:plmft11}), we have
\be
\cT_{\LnuB,{\rm local}}(\LnuB,t_0)\equiv
\int_0^\infty n_\LnuB(\LnuB,t)~\d t
=\int C_B P(C_B|L\bol) \cT_{L\bol,{\rm local}}(L\bol,t_0)~\d C_B.
\label{eq:TLnuBlocal}
\ee
Similar to equation (\ref{eq:Trelation}), we have
\be
\cT_{\LnuB,{\rm QSO}}(\LnuB,t_0)=\cT_{\LnuB,{\rm local}}(\LnuB,t_0);
\ee
and according to equation (\ref{eq:TMBQSO}), we have  
\be
\cT\MBQSO\MBpr=\cT\MBlocal\MBpr.
\ee
Considering that QSOs in the equation above include both optically bright
QSOs and obscured QSOs, we have
\be
\cT\MBQSOopt\MBpr\le\cT\MBlocal\MBpr.
\label{eq:TMBrelation}
\ee
The ratios of the time integral of the QSOLF and that predicted
from the local BHMF are given as follows:
\be
\cR\MBpr\equiv \frac{\cT\MBlocal\MBpr}
{\cT\MBQSO\MBpr}=1
\label{eq:R}
\ee
and
\be
\cR\opt\MBpr\equiv \frac{\cT\MBlocal\MBpr}
{\cT\MBQSOopt\MBpr}\ge 1.
\label{eq:Ropt}
\ee
The value of $\cR\opt\MBpr-1$ gives the number ratio of obscured QSOs to
optically bright QSOs.

\citet{Elvis94} study the spectral energy density distribution of a sample of
47 QSOs and find that the bolometric corrections in the $B$ band have a mean
value $\overline{C}_B=11.8$ with standard deviation $\Delta_{C_B}=4.3$.
The origin of the non-uniform of the bolometric correction (or the spectral
energy distribution) is not yet clear, which could be due to the difference
in the accretion rate, the BH mass and spin, etc.
For example, with increasing BH mass, the bolometric correction in the $B$
band may decrease since the peak of its spectral energy distribution may
move from a short wavelength toward the $B$ band or long wavelength,
and with increasing accretion rate, the bolometric correction in the $B$ band
may decrease since it may have a softer spectrum in the X-ray band.
In this paper, for simplicity, we assume that the probability distribution
function of the bolometric correction $P(C_B|L\bol)$ is independent of $L\bol$
and follows a Gaussian distribution as follows:
\be
P(C_B|L\bol)=\frac{1}{\sqrt{2\pi}\Delta_{C_B}}\exp\left[
-\frac{(C_B-\overline{C}_B)^2}{2\Delta^2_{C_B}}\right].
\label{eq:bolc2}
\ee
In the study below we use two values of $\Delta_{C_B}=4.3$ and 0
[$P(C_B|L\bol)=\delta(C_B-\overline{C}_B)$ if $\Delta_{C_B}=0$]
to check the effect of uncertainties in the distribution of the
bolometric corrections.

Note that inequalities (\ref{eq:TMBrelation}) and (\ref{eq:Ropt}) can be
generalized to other bands (e.g.,\ the X-ray band).
Given the observed QSOLF and the bolometric correction
in some other bands, it is worthy to make similar comparisons 
as is done in the B band below (see also discussion in
\S~\ref{sec:obscuration}).

\subsection{Observational constraints on the nuclear/QSO luminosity evolution
and BH growth}\label{sec:constraint}
\noindent
In \S~\ref{sec:Ltauage}, the model of the nuclear/QSO luminosity evolution is
characterized by three parameters: the period of the nuclear activity
in the first phase $\tau\I$,
the characteristic luminosity declining timescale in the second phase $\tau\D$,
and the efficiency $\epsilon$.
In \S~\ref{sec:parameters}, using inequality
(\ref{eq:TMBrelation}) or (\ref{eq:Ropt}) (see also eq.~\ref{eq:Trelation}),
we compare $\cT\MBQSOopt\MBpr$ with $\cT\MBlocal\MBpr$
obtained with different values of $\tau\I/\tau\Sp$, $\tau\D/\tau\Sp$, and
$\epsilon$ to provide observational constraints on the QSO model
and BH growth. 
In the figures of this section, the results are shown in the luminosity range
$-20>M_B>-30$. However, the results are fairly secure probably only in the
range $-23\ga M_B\ga -26$ \citep{Boyle00} (which are marked by two vertical
thin dot-dashed lines in the figures), and the results at $M_B\la -26$
might be affected by the uncertainty of the small number statistics.
In \S~\ref{sec:scatters}, we show the effects of the uncertainty of
the intrinsic scatter of the BH mass--velocity dispersion relation of local
galaxies and the scatter of the bolometric correction of QSOs
($\scatter$ and $\Delta_{C_B}$).

\subsubsection{$\tau\I/\tau\Sp$, $\tau\D/\tau\Sp$, $\epsilon$, and the
obscuration ratio}
\label{sec:parameters}
\noindent
In this subsection we always set $\scatter=0.27\dex$ \citep{Tremaine02} and
$\Delta_{C_B}=4.3$ \citep{Elvis94}.
With the luminosity evolution described in \S~\ref{sec:Ltauage},
below we study three models:
in model (a), only the first (or ``demand limited'')
phase is considered; in model (b), only the second (or supply limited)
phase is considered;
and in model (c), both the first and the second phases are considered.

\begin{figure}
\epsscale{0.5}
\plotone{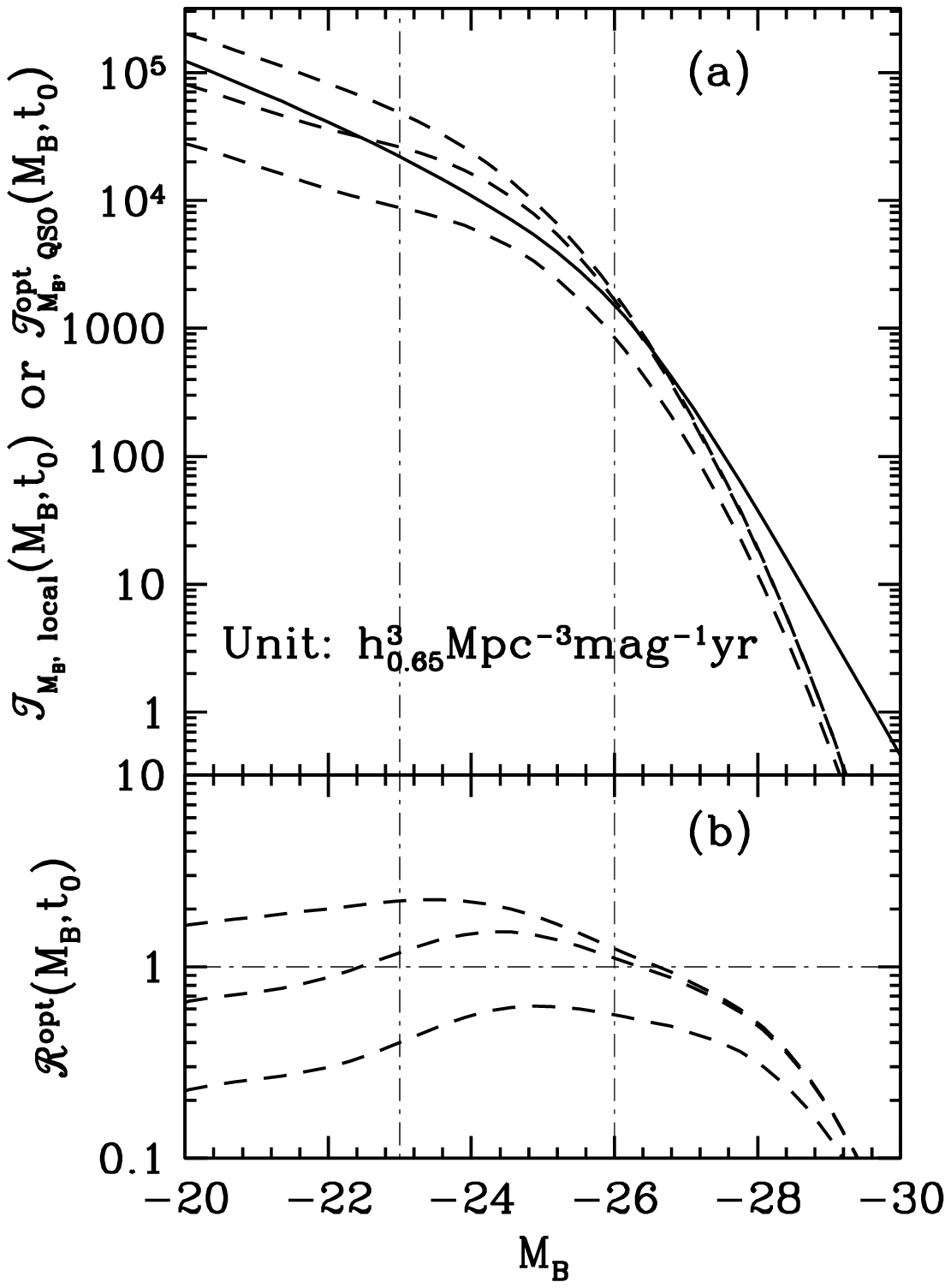}
\caption{Time integral of the observed LF of optically bright QSOs and
the time integral of the QSOLF predicted from local BHs.
The model of the luminosity evolution of the nuclear activity in
\S~\ref{sec:Ltauage} is used.
The curves are obtained by setting the parameters 
$\tau\D=0$, $\epsilon=0.1$, $\scatter=0.27\dex$, and $\Delta_{C_B}=4.3$.
(a) Dashed lines represent the time integral of the QSOLF
predicted from local BHs $\cT\MBlocal\MBpr$ (eqs.~\ref{eq:TLlocal},
\ref{eq:tauQSO}, \ref{eq:PLMQSO} and \ref{eq:TLnuBlocal});
solid line represents the time integral of the observed LF
of optically bright QSOs $\cT\MBQSOopt\MBpr$
(eqs.~\ref{eq:QSOLF} and \ref{eq:TMBQSO}).
(b) Ratio of the time integrals $\cR\opt\MBpr$ (eq.~\ref{eq:Ropt}).
The horizontal thin dot-dashed line represents $\cR\opt\MBpr=1$.
The two vertical thin dot-dashed lines mark the region $-23<M_B<-26$ in
which the LFs of optically bright QSOs are fairly secure \citep{Boyle00}.
In each panel, the dashed lines represent the results obtained with
$\tau\I/\tau\Sp=0.3, 1$, and $4$ from bottom to top at the faint end, respectively.
Inequality (\ref{eq:Ropt}) implies that the dashed line should be consistent
with the solid line in (a) or approximately equal to 1
in (b) if obscured QSOs are significantly less than optically bright QSOs,
and the dashed line should be higher than the solid line in (a) or
higher than 1 in (b)
if obscured QSOs are significantly numerous compared to optically bright QSOs.
To satisfy inequality (\ref{eq:Ropt}), the QSO lifetime $\tau\I$ is required
to be $\ga\tau\Sp\simeq 5\times 10^7\yr$ (if $\epsilon\simeq 0.1$).
See more discussion in \S~\ref{sec:parameters}.
}
\label{fig:model1sta}
\end{figure}

\begin{figure}
\epsscale{0.6}
\plotone{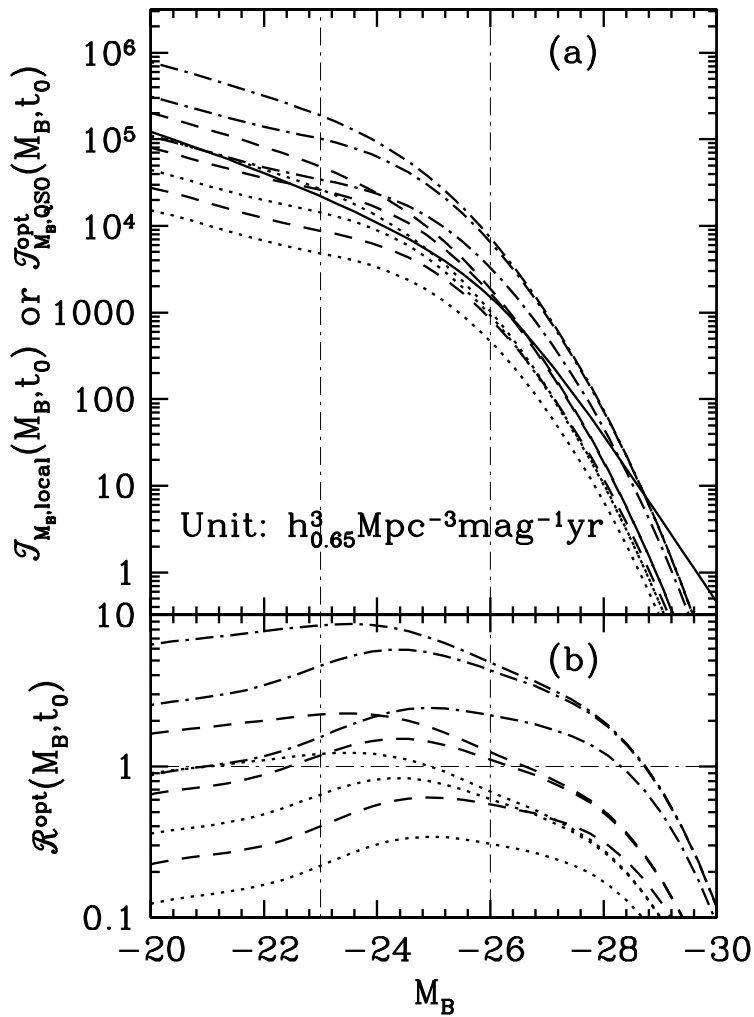}
\caption{Same as Figure~\ref{fig:model1sta}, except that there are
additional results obtained with $\epsilon=0.057$ and 0.31 (not only
$\epsilon=0.1$ in Fig.~\ref{fig:model1sta}) shown
as the dotted and dot-dashed lines, respectively.
The curves shift upward with increasing $\epsilon$.
As seen from (b), generally or at least at $M_B<-25$, $\cR\opt\MBpr$
obtained with $\epsilon=0.057$ (dotted lines) are smaller than 1,
which suggests that $\epsilon$ cannot be significantly smaller than 0.1.
See more discussion in \S~\ref{sec:parameters}.
}
\label{fig:eff}
\end{figure}

In model (a), $\tau\D=0$ and $\tau\I$ is a free parameter
(see eq.~\ref{eq:Lphase1}).
We first show the result obtained by setting $\epsilon=0.1$ in
Figure~\ref{fig:model1sta} and then see the change of the result by changing
the value of $\epsilon$ in Figure~\ref{fig:eff}.
\begin{itemize}
\item
In Figure~\ref{fig:model1sta}a,
$\cT\MBlocal\MBpr$ is shown as the dashed lines, 
and $\cT\MBQSOopt\MBpr$ is shown as the solid line;
and their ratios $\cR\opt\MBpr$ are shown as the dashed lines in panel (b).
As seen from Figure~\ref{fig:model1sta}(a) and (b),
$\cT\MBlocal\MBpr$ and $\cR\opt\MBpr$ increase with
increasing $\tau\I/\tau\Sp$ ($=0.3,1,4$ from bottom to top at the faint
end, respectively).
The reason of the increasing is that increasing $\tau\I/\tau\Sp$
will increase the mass range (or the upper limit of the mass range) of local
BHs that had nuclear luminosity $L$ in their
evolution history (see eqs.~\ref{eq:TLlocal} and \ref{eq:fI}).
In this model, we get the following constraints:

\renewcommand{\labelenumi}{(\roman{enumi})}
\begin{enumerate}
\item Constraints on $\tau\I/\tau\Sp$ from the sensitivity of
$\cT\MBlocal\MBpr$ to $\tau\I/\tau\Sp$:
as seen from Figure~\ref{fig:model1sta}(a), if $\tau\I/\tau\Sp\ga 1$, 
$\cT\MBlocal\MBpr$ becomes insensitive to the value of
$\tau\I/\tau\Sp$ at $M_B\la -26$.
Our calculations also show that if $\tau\I/\tau\Sp\ga 4$, $\cT\MBlocal\MBpr$
becomes insensitive to the values of $\tau\I/\tau\Sp$ at $M_B\la -23$.
For example, we have $\cT\MBlocal(\tau\I/\tau\Sp=10)
/\cT\MBlocal(\tau\I/\tau\Sp=1)-1\la 10\%$ at $M_B\la -26$ and 
$\cT\MBlocal(\tau\I/\tau\Sp=10)/\cT\MBlocal(\tau\I/\tau\Sp=4)-1\la 15\%$
at $M_B\la -23$.
The reason for this insensitivity is that the additional contribution to
$\cT\MBlocal\MBpr$ at a given $M_B$ due to the increase of $\tau\I/\tau\Sp$
comes from the BHs at the high-mass end (see eqs.~\ref{eq:TLlocal} and
\ref{eq:fI}).
Thus, for a given $M_B$, if $\tau\I/\tau\Sp$ is larger than a certain value,
the masses of the BHs from which the additional contribution comes will be
significantly high, and the additional contribution may be negligible
since the local BHMF decreases sharply (or exponentially) at the high-mass end
($M\bhpr\ga 3\times 10^8\msun$).
The insensitivity of $\cT\MBlocal\MBpr$ to the value of
$\tau\I/\tau\Sp$ above suggests that if the real $\tau\I/\tau\Sp$ is larger
than 4, it is more likely that only the lower limit of $\tau\I/\tau\Sp$
($\sim 4$ here), rather than an accurate value of $\tau\I/\tau\Sp$, can be
provided by fitting $\cT\MBlocal\MBpr$ to $\cT\MBQSO\MBpr$.
To get an accurate estimate of $\tau\I/\tau\Sp$, it would require either
observations at fainter luminosities or precise
measurements of both the QSOLF and the local BHMF
(e.g.,\ with error much less than 10\%) within the current luminosity range.
The insensitivity of $\cT\MBlocal$ to $\tau\I/\tau\Sp$ still holds
if $\tau\D$ is non-zero (see eqs.~\ref{eq:Mbhpr} and \ref{eq:fI}).
The suggestion above will also not be affected by changing the value of
$\epsilon$,
since as will be seen below (Fig.~\ref{fig:eff}), changing $\epsilon$ does not
affect the shape of $\cT\MBlocal\MBpr$ or $\cR\opt\MBpr$ for a given
$\tau\I/\tau\Sp$.

\item Constraints on $\tau\I/\tau\Sp$ from the value of $\cR\opt\MBpr$:
Figure~\ref{fig:model1sta}(b) shows that $\cR\opt\MBpr$ is
significantly smaller than 1 if $\tau\I/\tau\Sp\ll 1$ (see the bottom dashed
line with $\tau\I/\tau\Sp=0.3$);
and $\cR\opt\MBpr$ can roughly exceed or be around 1
at $-23\ga M_B\ga -26$ if $\tau\I/\tau\Sp\ga 1$ (see the middle and top
dashed lines obtained with $\tau\I/\tau\Sp=1$ and 4).
Thus, to satisfy inequality (\ref{eq:Ropt}), the QSO lifetime should
be $\ga \tau\Sp\simeq5\times 10^7\yr$ if $\epsilon\simeq0.1$.
This conclusion will not be affected after setting a non-zero timescale
$\tau\D$, since as will be seen below (Fig.~\ref{fig:model3}),
increasing $\tau\D$ decreases $\cR\opt\MBpr$ at least at $M_B\sim -26$.

\item Constraints on the ratio of obscuration:
as seen from Figure~\ref{fig:model1sta}b,
if $\tau\I/\tau\Sp \ga 4$, $\cR\opt\MBpr$ is about 2 at $M_B\simeq -23$
and about 1 at $M_B\simeq -26$.
This result suggests that if $\tau\I/\tau\Sp \ga 4$,
there should exist some obscured QSOs (which are missed by optical surveys),
and the ratio of obscured QSOs to unobscured (or optically bright) QSOs
may be about 1 at the faint end (intrinsic absolute
magnitude $M_B\sim -23$);
compared to optically bright QSOs, few obscured QSOs are expected to
exist at the luminous end ($M_B\sim -26$). 
If $\tau\I/\tau\Sp\simeq 1$, few obscured QSOs/AGNs are expected to exist
at both faint and bright ends.
Note that these conclusions are obtained with $\epsilon=0.1$
and $\tau\D=0$, and may be affected if $\epsilon\ga 0.1$ or $\tau\D>0$.
\end{enumerate}

\item In Figure~\ref{fig:eff}a and 4b, the dotted, dashed, and dot-dashed
lines represent the results obtained with $\epsilon=0.057$ (the efficiency for
the thin-disk accretion onto a Schwarzschild BH), 0.1, and 0.31 (the maximum
efficiency allowed for the thin-disk accretion onto a Kerr BH with
dimensionless spin parameter 0.998, see \citealt{Thorne74};
here we do not consider the possibility of extremely high efficiency, see, e.g.,
\citealt{NIA03}; \citealt{LP00}), respectively.
As in Figure~\ref{fig:model1sta}, for each type of those lines, we show the
results obtained with $\tau\I/\tau\Sp=4,1$, and $0.3$ (top, middle, and bottom
lines, respectively).
As seen from Figure~\ref{fig:eff}, $\cT\MBlocal\MBpr$ and $\cR\opt\MBpr$
increase with increasing $\epsilon$ without changing the shape
(which can also be inferred from eqs.~\ref{eq:tauSp} and \ref{eq:PLMQSO}).
We obtain the following constraints from Figure~\ref{fig:eff}.

\begin{enumerate}
\item Constraints on $\epsilon$:
as seen from Figure~\ref{fig:eff}(b), if $\epsilon=0.057$ (dotted lines),
$\cR\opt\MBpr$ are generally smaller than 1,
which suggests that the efficiency of QSOs/AGNs cannot 
be significantly less than 0.1.

\item Constraints on $\tau\I/\tau\Sp$:
our calculations show that if $\epsilon=0.31$,
to satisfy inequality (\ref{eq:Ropt}),
it is required that $\tau\I/\tau\Sp\ga 0.2$ or the QSO lifetime
should be $\ga 4\times 10^7\yr$
(see also the bottom dot-dashed line in Fig.~\ref{fig:eff}b).

\item Constraints on the ratio of obscuration:
as seen from Figure~\ref{fig:eff}(b), if $\epsilon=0.31$ and
$\tau\I/\tau\Sp\ga 4$ (see the top dot-dashed line), 
$\cR\opt\MBpr$ is $\sim 8$ at $M_B\sim -23$ and $\sim 4 $ at $M_B\sim -26$,
which suggests that the upper limit of the ratio of obscured QSOs/AGNs to
un-obscured QSOs/AGNs is $\sim 7$ at $M_B\sim -23 $ and $\sim 3$
at $M_B\sim -26$.
\end{enumerate}

The constraints obtained above from Figure~\ref{fig:eff} will not be affected
by having a non-zero $\tau\D$,
since as will be seen below (Fig.~\ref{fig:model3}),
increasing $\tau\D$ decreases the $\cT\MBlocal\MBpr$ or $\cR\opt\MBpr$
at least at $M_B\sim -26$.

\end{itemize}

\begin{figure}
\epsscale{0.6}
\plotone{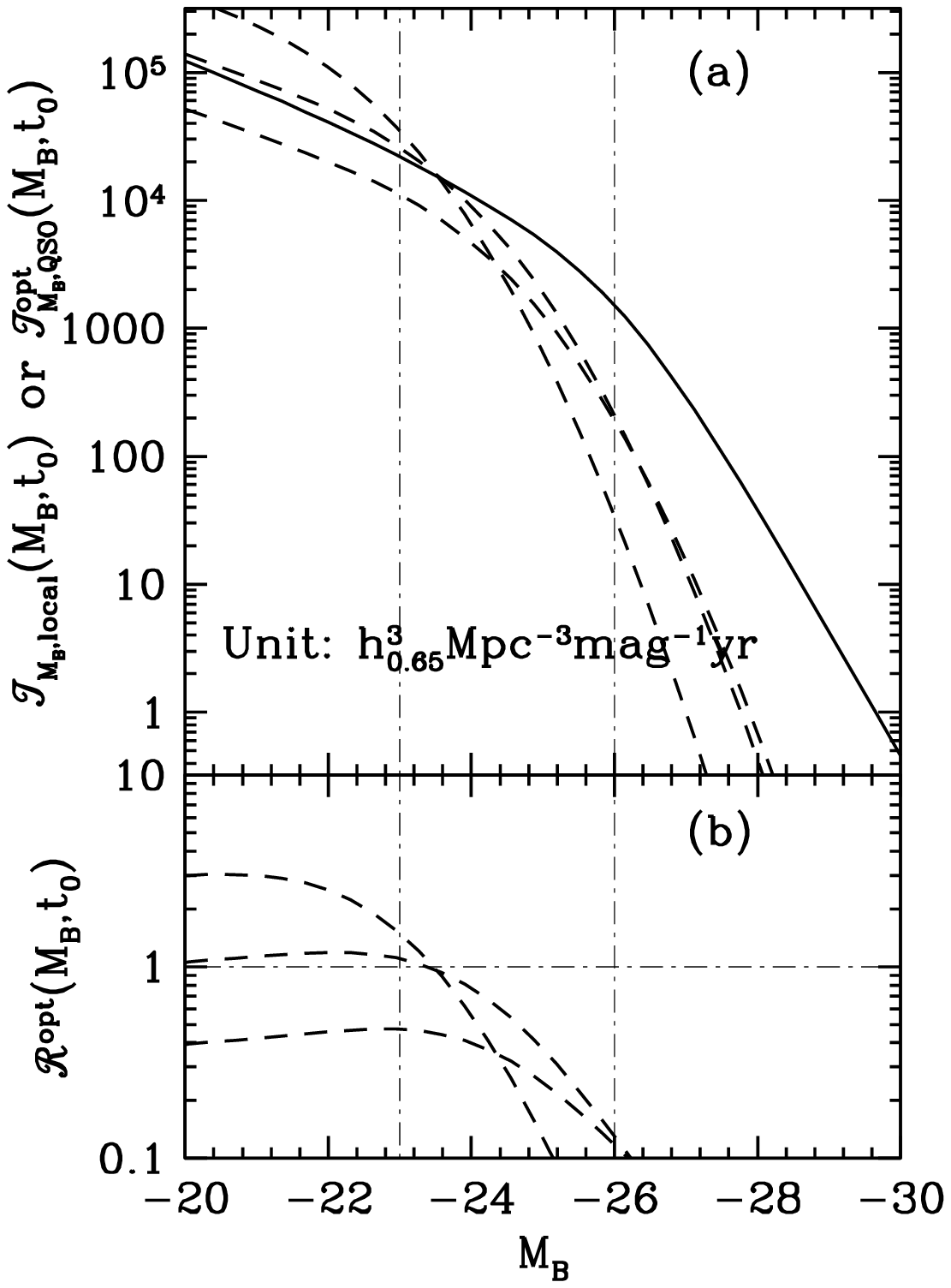}
\caption{Same as Figure~\ref{fig:model1sta}, but with $\tau\I=0$ and
$\tau\D/\tau\Sp=4, 1$, and $0.3$ (from top to bottom at the faint end, respectively;
dashed lines; model [b]).
(b) $\cR\opt(M_B\sim-26)$ is significantly smaller than 1
(which is still true even if $\epsilon=0.31$).
This figure suggests that to satisfy inequalities (\ref{eq:TMBrelation}) and
(\ref{eq:Ropt}), it is impossible to only have the
luminosity-declining phase (eq.~\ref{eq:Lphase2}, the second phase described
in \S~\ref{sec:Ltauage}) in the QSO luminosity evolution.
See \S~\ref{sec:parameters}.
}
\label{fig:model2}
\end{figure}

\begin{figure}
\epsscale{0.6}
\plotone{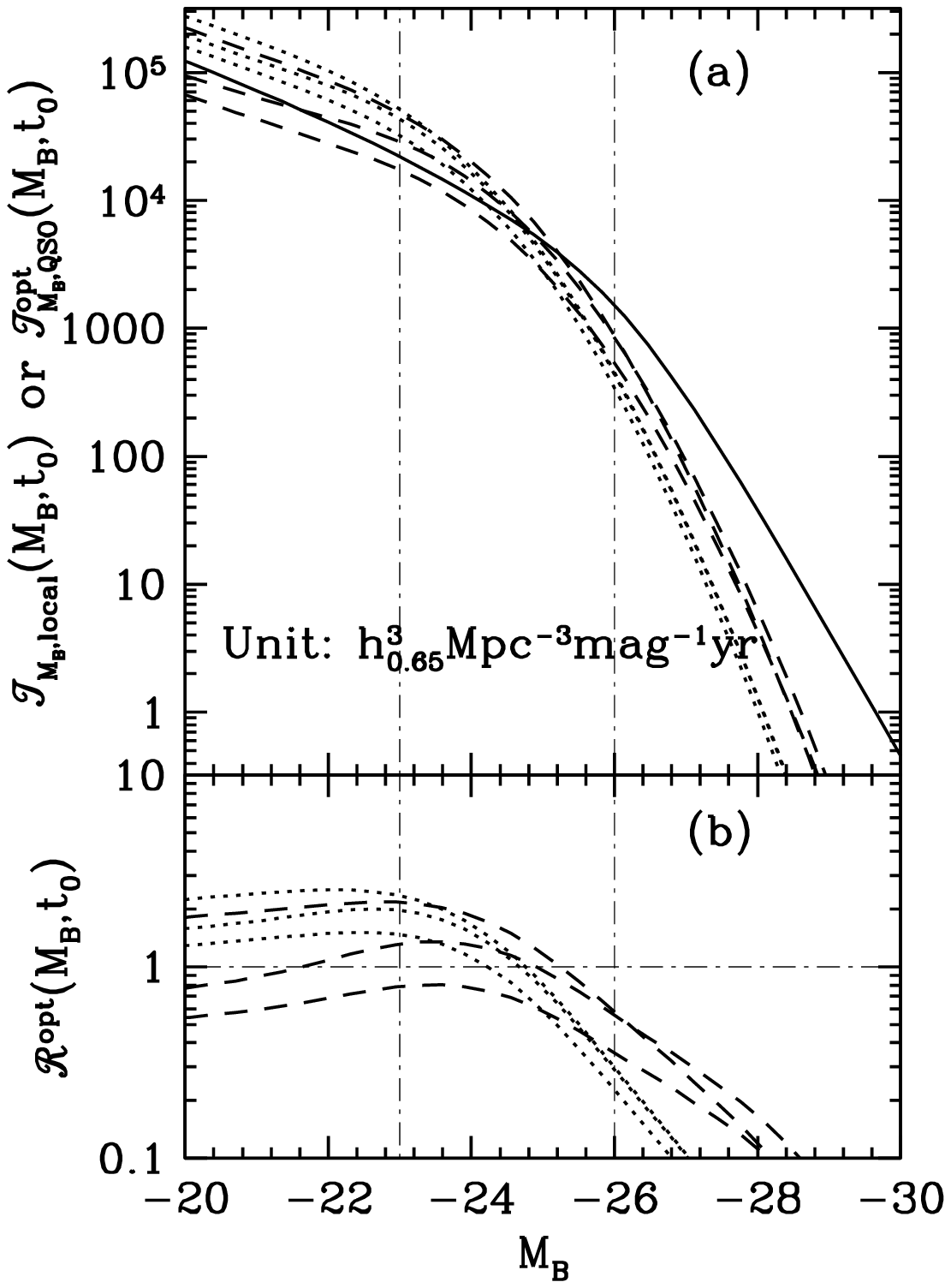}
\caption{Same as Figure~\ref{fig:model1sta}, but with $\tau\D/\tau\Sp=0.3$
(dashed lines) and 1 (dotted lines; model [c]).
(b) $\cR\opt(M_B\sim-26)$ is significantly smaller than 1,
which suggests that to satisfy inequalities (\ref{eq:TMBrelation}) and
(\ref{eq:Ropt}), $\tau\D$ should be significantly shorter than $\tau\Sp$
(e.g.,\ $\tau\D<0.3\tau\Sp$), or the luminosity-declining phase (the second
phase described in \S~\ref{sec:Ltauage})
should not dominate the growth of local BHs.
See \S~\ref{sec:parameters}.
}
\label{fig:model3}
\end{figure}

In model (b), $\tau\I=0$ and $\tau\D$ is a free parameter (see
eq.~\ref{eq:Lphase2}).
Setting $\epsilon=0.1$, the results obtained with $\tau\D/\tau\Sp=0.3,1$, and $4$
are shown as dashed lines from bottom to top at the faint end, respectively,
in Figure~\ref{fig:model2}.
As seen from Figure~\ref{fig:model2}, for any given $\tau\D$,
$\cT\MBlocal\MBpr$ is smaller than $\cT\MBQSOopt\MBpr$ at $M_B\la -26$
at least by a factor of $\sim 10$ if $\epsilon=0.1$ 
(and at least by a factor of 2 or 3 if $\epsilon=0.31$).
This result suggests that to satisfy the expected relation between the QSOLF
and local BHMF (inequalities~\ref{eq:TMBrelation} or \ref{eq:Ropt}),
it is impossible to only have the luminosity-declining phase in the QSO
luminosity evolution.

In model (c), both $\tau\I$ and $\tau\D$ are not zero.
Setting $\tau\I/\tau\Sp=4,1,0.3$ and $\epsilon=0.1$,
we show the results obtained with $\tau\D/\tau\Sp=0.3$ and 1 as the dashed and
dotted lines, respectively, in Figure~\ref{fig:model3}.
The value of $\tau\D$ affects the shape and value of $\cT\MBlocal\MBpr$
and $\cR\opt\MBpr$.
As seen from Figures~\ref{fig:model1sta} and \ref{fig:model3},
increasing $\tau\D/\tau\Sp$ decreases $\cT\MBlocal\MBpr$ and $\cR\opt\MBpr$
at high luminosities (e.g.,\ $M_B\la -26$).
Figure~\ref{fig:model3}b shows that at $M_B\sim-26$, $\cR\opt\MBpr$ is
significantly smaller than 1 (about 0.2--0.3 if $\tau\D/\tau\Sp=1$
and about 0.3--0.6 if $\tau\D/\tau\Sp=0.3$),
which suggests that $\tau\D$ should be significantly shorter than $\tau\Sp$,
or the second (or supply limited) phase should not dominate the growth of
local BHs (see eq.~\ref{eq:Mbhpr}).
Our calculation shows that this suggestion will not be significantly
affected even if $\epsilon=0.31$.

By combining the constraints obtained in the three models above, below we
summarize the results on the fundamental parameters of the QSO luminosity
evolution and BH growth:

\begin{enumerate} 
\item The QSO mass-to-energy conversion efficiency is
$\epsilon\ga 0.1$ (see Fig.~\ref{fig:eff}).

\item The period of the nuclear activity 
is longer than $\sim\tau\Sp\simeq 5\times 10^7\yr$ if $\epsilon=0.1$
and longer than $\sim 0.2\tau\Sp=4\times 10^7\yr$ if $\epsilon=0.31$
(see Figs.~\ref{fig:model1sta} and \ref{fig:eff}).

\item If the real $\tau\I$ is larger than a certain value ($\sim 4\tau\Sp$
here), it is difficult to provide an accurate estimate on the value of $\tau\I$
unless observations are extended to fainter luminosities or
precise measurements of both the QSOLF and the local BHMF within the current
luminosity range are available (e.g.,\ with error much less than 10\%;
see Fig.~\ref{fig:model1sta}).

\item The characteristic declining timescale of the luminosity evolution in
the second phase $\tau\D$ should be significantly shorter than $\tau\Sp$
(e.g.,\ $\tau\D<0.3\tau\Sp$ if $\epsilon=0.1$ and $\tau\D<\tau\Sp$
if $\epsilon=0.31$; see Fig.~\ref{fig:model3}); thus, the second phase
does not dominate the growth of local BHs (see eq.~\ref{eq:Mbhpr}).

\item There might exist a large number of obscured QSOs/AGNs, but the ratio of 
the obscured QSOs/AGNs to the unobscured (or optically bright) QSOs should
be not larger than 7 at $M_B\sim -23$ and 3 at $M_B \sim -26$
if $\epsilon\simeq 0.31$ (see the top dotted line in Fig.~\ref{fig:eff})
and not larger than 1 at $M_B\sim -23$ and negligible at $M_B\sim -26$
if $\epsilon=0.1$ (see Fig.~\ref{fig:model1sta}).
\end{enumerate}

According to the results above, the lower limit of the QSO lifetime (defined
directly through the luminosity evolution of individual QSOs) is $\simeq
4\times 10^7\yr$.  There are also other independent methods (e.g.,\
\citealt{Martini03}), such as the clustering of QSOs \citep{MW01,HH01} or the
modeling of the QSOLF in hierarchical galaxy formation scenario
\citep{KH00,HL98}, to provide constraints on the QSO lifetime (usually in the
range of $10^6-10^8\yr$). Here we do not make a comparison with those results
since the meanings of the QSO lifetime in different methods or contexts are not
exactly the same. The upper limit of the lifetime can be usually constrained by
the rising and falling of the characteristic luminosity of the entire QSO
population as a function of cosmic time (roughly a few times $10^9\yr$; see
eqs.~\ref{eq:Mstarzpei} and \ref{eq:Mstarz}).

It is worthy to note that item 3 above would hold for many other methods to
estimate the QSO lifetime by using the QSOLF, not only for the specific
study in this paper, as a result of the sharp decrease of the QSOLF at the
bright end and the limited luminosity range of the observations.

Item 5 above does not exclude the possibility of the existence of a large
number of obscured QSOs at high luminosities if the efficiency is high
(e.g.,\ $\epsilon\sim 0.3$), which is slightly different from the result
obtained by comparing partial mass densities (see eq.~\ref{eq:YT}) in
\citet{YT02} that obscured accretion is not important for the growth of
high-mass BHs ($>10^8\msun$).
The reason for the difference is partly because, in the present study, we adopt
a high value of $\scatter$($=0.27\dex$) and include the effect of the scatter
of the bolometric corrections (for the detailed effects of $\scatter$ and
$\Delta_{C_B}$ see \S~\ref{sec:scatters} and Figs.~\ref{fig:model1sig} and
\ref{fig:model1bol} below).
Furthermore, \citet{YT02} compare the partial mass density accreted in QSOs
having luminosity higher than a certain value with the partial mass density
in local BHs (see eq.~\ref{eq:YT}), and the present study compares the number
density (or the time-integral of the number density) at a given luminosity;
hence, the result of the obscuration obtained in \citet{YT02} is mainly for the
ratio of all the BH {\em mass} accreted in obscured QSOs with intrinsic
luminosity higher than a certain value,
rather than the {\em number} ratio of obscured QSOs at a given luminosity
in the present study.  

In addition, in Figures~\ref{fig:model1sta}--\ref{fig:model3}
($\epsilon\la 0.31$),
inequalities (\ref{eq:TMBrelation}) and (\ref{eq:Ropt}) are always
not satisfied at the bright end $M_B\la-29$ --- $-26$, which is mainly because
the local BHMF declines more sharply than the QSOLF at the bright end
(note that the velocity dispersion function of early-type galaxies is
fitted by an exponential form at the bright end, but the QSOLF is fitted by
a power law).
The forms of the QSOLF and local BHMF at the bright end are probably
affected by the uncertainty of the small number statistics;
otherwise, the physical mechanism/parameters (e.g.,\ QSO efficiency) of very
luminous QSOs ($M_B\la-29$ --- $-26$) or the properties of nearby galaxies
with very big BHs ($\ga 10^9\msun$) should be very different from those
of the main population of QSOs or nearby early-type galaxies.

\subsubsection{Is it possible that most QSOs radiate at super-Eddington
luminosities?}\label{sec:superEdd}
\noindent
In \S~\ref{sec:Ltauage} it is assumed that after the nuclear activity is
triggered, QSOs first radiate at the Eddington luminosity for a period
$\tau\I$ and then radiate at sub-Eddington luminosities as a result of the
decline of available accretion material supply.
However, accretion with super-Eddington luminosities might occur in some cases.
For example, when BH mass is small and the accretion rate is sufficiently
high, the outflow pushed by the radiation may be trapped by the infalling
gas and the energy is radiated away at a rate higher than the Eddington limit
(see discussion in \citealt{Blandford03}),
or when some strong density inhomogeneity is developed in a thin disk,
the disk may also radiate at super-Eddington luminosity \citep{Begelman02}.

\begin{figure}
\epsscale{0.6}
\plotone{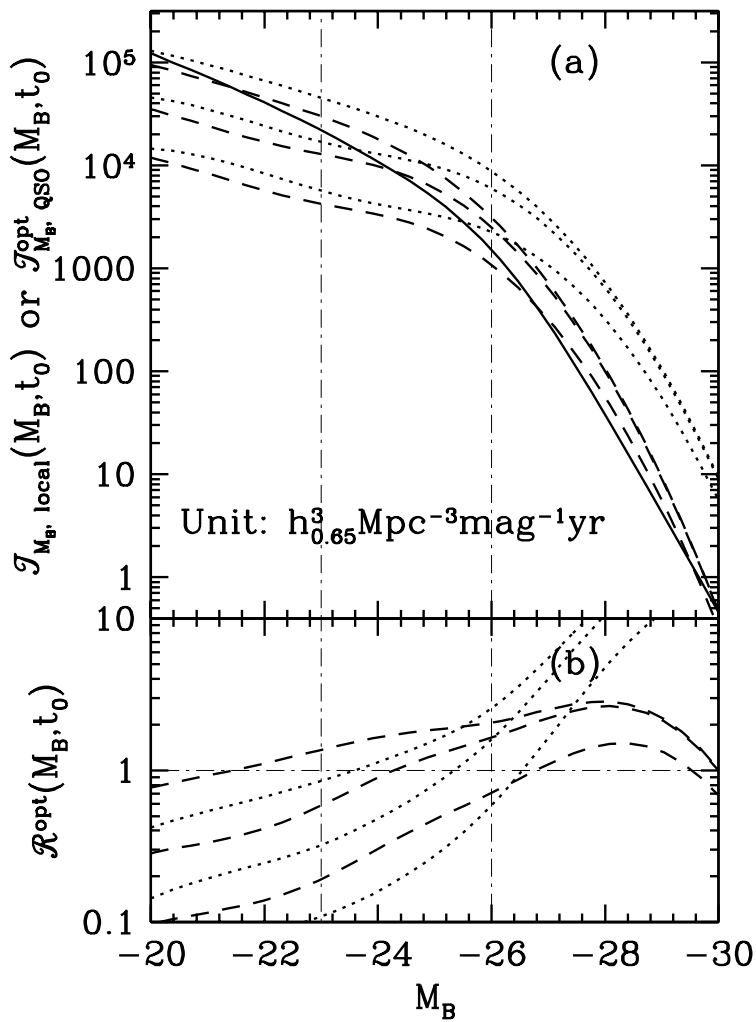}
\caption{Same as Figure~\ref{fig:model1sta}, except that the QSOs are assumed
to radiate at a luminosity higher than the Eddington luminosity
by a factor of $l$ ($\tau\Sp'=l^{-1}\tau\Sp$, $\tau\I/\tau\Sp'=4, 1, 0.3$,
$\tau\D=0$, and $\epsilon=0.1$).
The dashed lines show the results obtained with $l=2$,
and the dotted lines show the results obtained with $l=5$.
As seen from (b), at $-23>M_B>-26$, $\cR\opt\MBpr$ either is smaller
than 1, which is inconsistent with inequality (\ref{eq:Ropt}),
or increases with increasing luminosity, which is inconsistent with current
observations on obscured AGNs (e.g.,\ \citealt{Barger03}).
This figure suggests that it is unlikely that most QSOs radiate
at a luminosity much higher than the Eddington luminosity (or accrete at an
accretion rate much higher than the Eddington accretion rate).
See details in \S~\ref{sec:superEdd}.
}
\label{fig:supedd}
\end{figure}

Below we will see whether the expected relations given by equations (\ref{eq:TMBrelation})
and (\ref{eq:Ropt}) can be satisfied if all QSOs radiate at a luminosity
higher than the Eddington luminosity, say, by a factor of $l>1$.
Note that in this case the value of the characteristic increasing timescale of
the luminosity or BH mass, denoted by $\tau\Sp'$, is smaller than the value in
equation (\ref{eq:tauSp}) by a factor of $l$.
In Figure~\ref{fig:supedd} we show the time integral of the LFs
obtained with $\tau\I/\tau\Sp'=4, 1, 0.3$ (from top to bottom) and
with $l=2$ (dashed lines) and $l=5$ (dotted lines). 
The other parameters (such as $\tau\D=0$, $\epsilon=0.1$, etc.) in
Figure~\ref{fig:supedd} are the same as those in Figure~\ref{fig:model1sta}.
As seen from Figure~\ref{fig:supedd}(b), at $-23>M_B>-26$, either
$\cR\opt\MBpr$ is smaller than 1, which is inconsistent with inequality
(\ref{eq:Ropt}), or $\cR\opt\MBpr$ increases with increasing luminosity,
which is inconsistent with current observations that few type II QSOs
are observed and most observed obscured AGNs are at low luminosities
(e.g.,\ \citealt{Barger03}).
The results above suggest that it is unlikely that most QSOs radiate
at a luminosity much higher than the Eddington luminosity (or accrete at an
accretion rate much higher than the Eddington accretion rate).
We expect that accretion with super-Eddington luminosities operates
maybe only at the very early stage of the nuclear activity, and will not
significantly affect our results obtained in \S~\ref{sec:constraint}.

\subsubsection{Effects of the uncertainty of $\scatter$ and $\Delta_{C_B}$}
\label{sec:scatters}
\begin{figure}
\epsscale{0.6}
\plotone{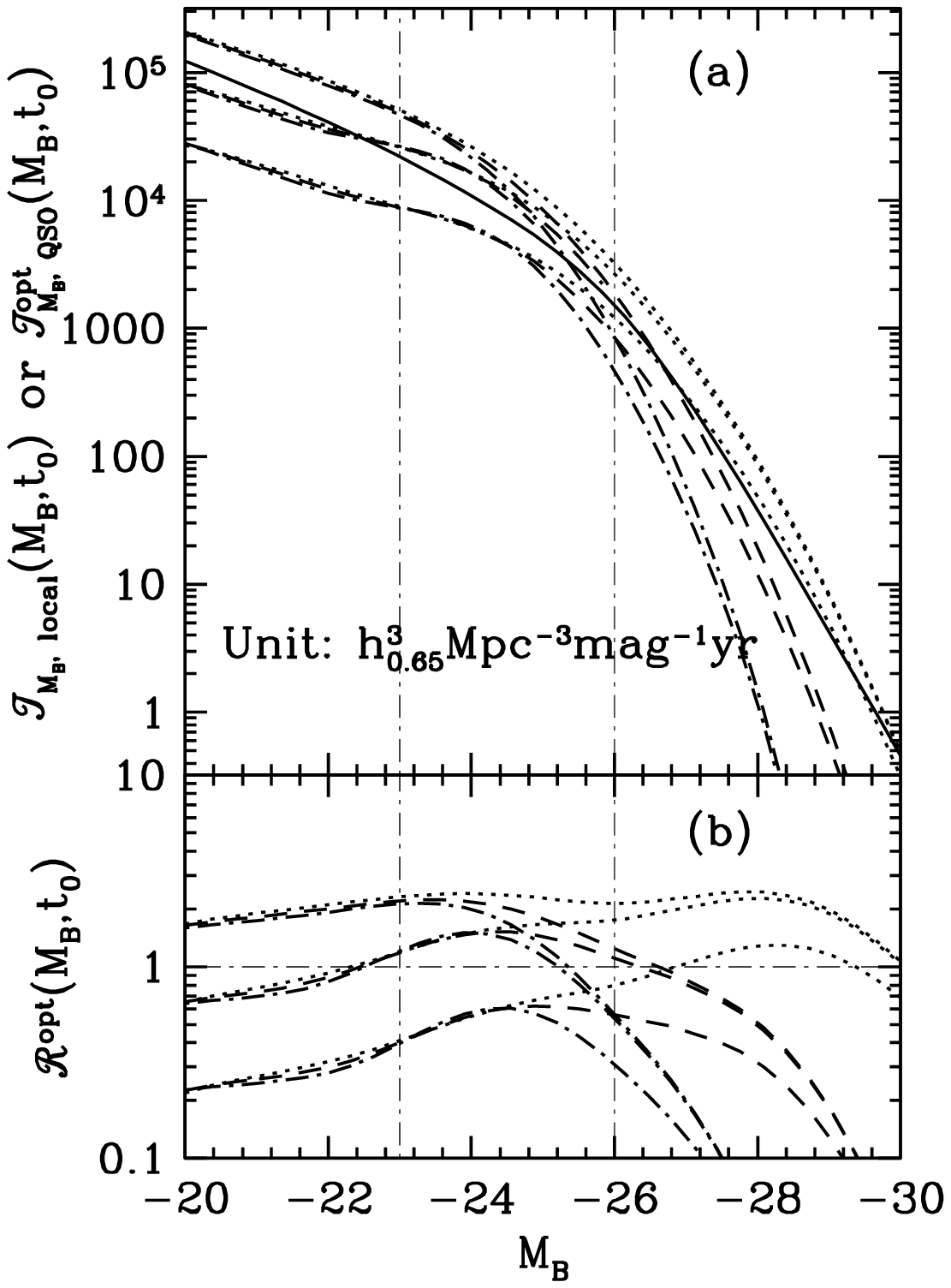}
\caption{Same as Figure~\ref{fig:model1sta}, except that
there are additional results obtained with $\scatter=0$ and $0.4\dex$
(not only $\scatter=0.27\dex$ in Fig.~\ref{fig:model1sta}) shown
as the dot-dashed and dotted lines, respectively.
The effect of the uncertainty of $\scatter$ is negligible at the faint end
($M_B\ga -24$) but may be significant at the bright end ($M_B\la -26$).
See \S~\ref{sec:scatters}.
}
\label{fig:model1sig}
\end{figure}

\noindent
The effect of the uncertainty of $\scatter$ is shown in
Figure~\ref{fig:model1sig}.
In Figure~\ref{fig:model1sig} we show $\cT\MBlocal\MBpr$ and $\cR\opt\MBpr$
obtained by setting $\scatter=0$ (dot-dashed lines) and $0.4\dex$ (dotted
lines), as well as the results obtained in Figure~\ref{fig:model1sta}
(shown by the dashed lines with $\scatter=0.27\dex$ and the solid line).
The other parameters of the dot-dashed and the dotted lines are the same
as in Figure~\ref{fig:model1sta}.
As seen from Figure~\ref{fig:model1sig}(b),
the effect of the uncertainty of $\scatter$ is negligible at $M_B\ga -24$;
however, $\cR\opt\MBpr$ or $\cT\MBlocal\MBpr$
is significantly affected by the value of $\scatter$ at $M_B\ga -26$.
For example, at $M_B=-26$, $\cR\opt\MBpr$ obtained with
$\scatter=0.4\dex$ (dotted lines) is larger than that obtained with
$\scatter=0.27\dex$ (dashed lines) by a factor of $\sim 2$
and is larger than that obtained with $\scatter=0$ (dot-dashed lines)
by a factor of more than $\sim 4$.

\begin{figure}
\epsscale{0.6}
\plotone{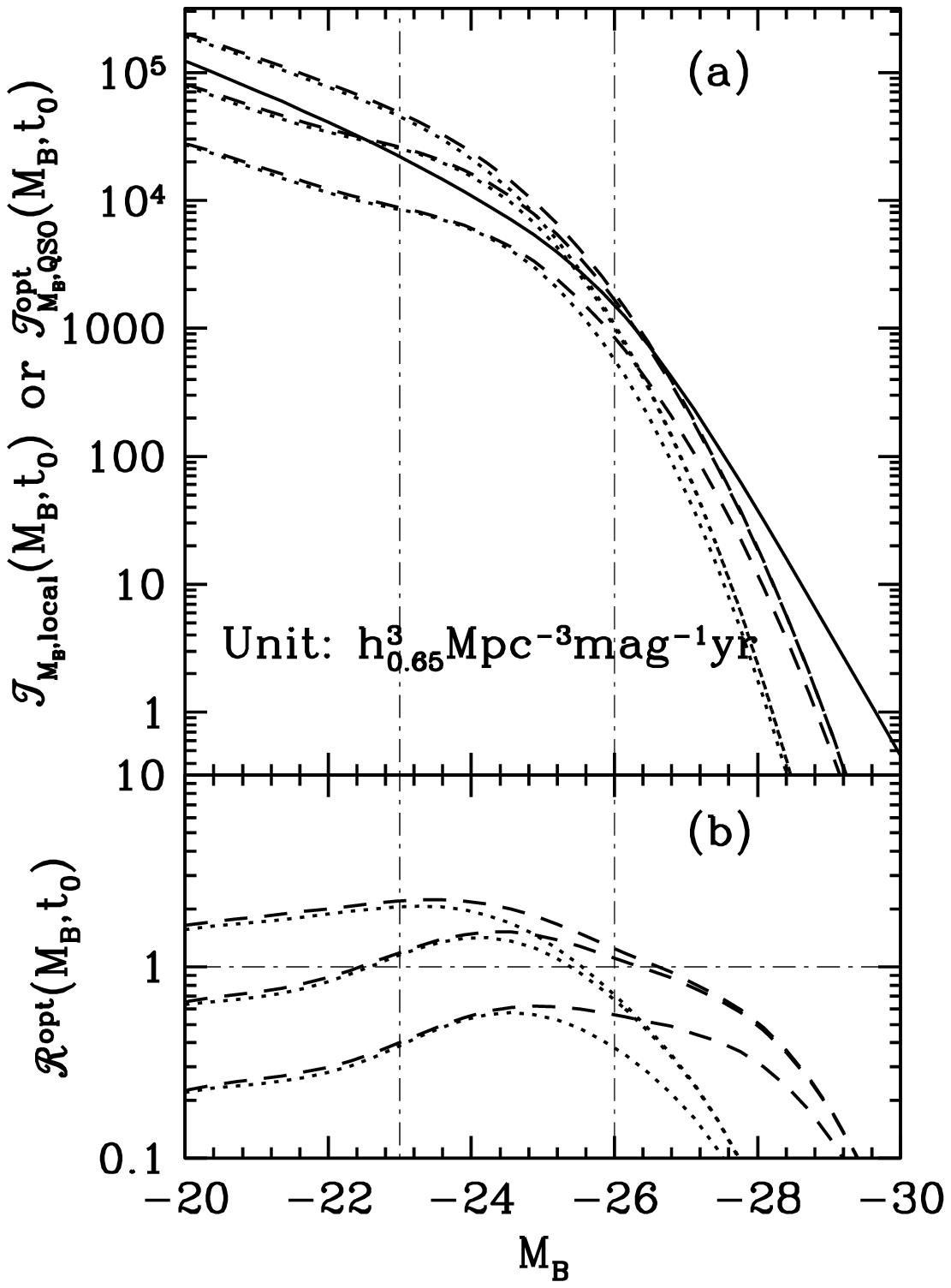}
\caption{Same as Figure~\ref{fig:model1sta}, except that there are additional
results obtained with $\Delta_{C_B}=0$ (not only $\Delta_{C_B}=4.3$
in Fig.~\ref{fig:model1sta}) shown as the dotted lines. 
The effect of the uncertainty of $\Delta_{C_B}$ is negligible at the faint end
($M_B\ga -24$) but may be significant at the bright end ($M_B\la -26$).
See \S~\ref{sec:scatters}.
}
\label{fig:model1bol}
\end{figure}

The effect of the uncertainty of $\Delta_{C_B}$ is shown in
Figure~\ref{fig:model1bol}.
In Figure~\ref{fig:model1bol} we show the results obtained by setting
$\Delta_{C_B}=0$ as the dotted lines, as well as those obtained
in Figure~\ref{fig:model1sta} (dashed lines; $\Delta_{C_B}=4.3$).
The other parameters of the dotted lines are the same as in
Figure~\ref{fig:model1sta}.
As seen from Figure~\ref{fig:model1bol}(b), at the faint end ($M_B\la-24$),
$\cR\opt\MBpr$ [or $\cT\MBlocal\MBpr$] is
insensitive to the value of $\Delta_{C_B}$;
however, at the luminous end ($M_B \la -26$), $\cR\opt\MBpr$ obtained
with $\Delta_{C_B}=4.3$ (dashed lines) is larger than that obtained with
$\Delta_{C_B}=0$ (dotted lines; e.g., by a factor of about 3 at $M_B=-27$).

The results above raise the importance of accurately measuring both
$\scatter$ and $\Delta_{C_B}$ and of studying the dependence of the bolometric
correction on the BH mass, the accretion rate, or other physical parameters
for precise understanding of the relation
between the local BHMF and the QSOLF
(especially at the luminous end, e.g.,\ $M_B\la -26$),
the luminosity evolution of the nuclear activity, and the BH growth.

\section{Discussions}\label{sec:discussion}
\subsection{The QSO luminosity evolution and the triggering history of the
accretion onto seed BHs}
\noindent
As part of the steps to understand the physics behind the QSO
phenomenon and BH growth, it is important to investigate both the luminosity
evolution and the triggering history of the accretion onto their seed BHs,
which together control the shape and the value of the QSOLF
(see eqs.~\ref{eq:nL} and \ref{eq:calN}).
In the currently popular coevolution scenario of QSOs and galaxy spheroids
in the cold dark matter (CDM) cosmology, it is generally assumed that QSOs are triggered by
hierarchical (major) mergers of galaxies, and the triggering rate is 
controlled by the (major) merger rate of galaxies (or halos in less fine models;
e.g.,\ \citealt{KH00,HL98} and references therein).
With an assumed luminosity evolution (e.g.,\ usually a step function or an
exponentially declining function in those models),
the coevolution models can reproduce the observed QSOLF. 
However, in those models, as a result of many uncertainties in the estimate of the
galaxy merger rate and the gas infalling rate for BH growth,
it is hard to differentiate whether the change of the QSOLF is due to a change
of the nuclear luminosity evolution or to a change of the triggering
rate; hence, it is hard to give an accurate constraint on the QSO
luminosity evolution.
For example the QSOLF can be reproduced for a large range of the assumed
lifetime ($10^6$ to $10^8\yr$; e.g.\ \citealt{KH00,HL98}).

In this paper, by investigating the relation between the QSOLF and the local
BHMF, we separate the luminosity evolution of individual QSOs from the
triggering rate of the QSO population, with the assumption that
each local massive BH has experienced the QSO phases and BH mergers are
ignored.
As shown in equations (\ref{eq:TLlocal}) and (\ref{eq:Trelation}), the
triggering rate is circumvented, and only the luminosity evolution of
individual
QSOs is implicitly reflected by their lifetime $\tau\life(M\bhpr)$ and the
luminosity probability distribution $P(L|M\bhpr)$ in their evolution history. 
Thus, the constraints on the QSO luminosity evolution (such as the QSO
lifetime and the efficiency) obtained here do not depend on the poorly known
QSO triggering history, and we expect that these obtained constraints could be
further used to infer the triggering history of seed BHs or
refine the coevolution model for QSOs and galaxy spheroids (see also
\citealt{Blandford03}).

\subsection{Obscured QSOs/AGNs}\label{sec:obscuration}
\noindent
Based on the assumption that the extragalactic X-ray background is
mainly contributed by QSOs/AGNs, a larger number of obscured QSOs/AGNs
(more than the optically bright or unobscured QSOs by a factor of $\ga 4$)
were expected to exist from the X-ray background synthesis model
\citep[e.g.,][and references therein]{GSH01}.
The existence of those sources is generally consistent with the expectation of
the unification model of QSOs/AGNs \citep{Antonucci93}: if the dusty torus is
along the line of sight, the light from these QSOs/AGNs will be absorbed by
the dusty torus (this structure is proposed to explain 
the classification of Seyfert 1 and Seyfert 2 galaxies at low redshift $z<1$)
and these QSOs/AGNs might be missed from observations in the optical band.
In this scenario, obscuration is generally a geometric effect, and the
obscured fraction is determined by the opening angle of the dusty torus.
If the opening angle is the same for all QSOs/AGNs,
the fraction of obscured QSOs/AGNs will be independent of the intrinsic
luminosity of QSOs/AGNs.

Recent X-ray observations by Chandra and XMM-Newton confirm the existence
of obscured AGNs/QSOs (e.g.,\ \citealt{Barger03}).
However, the fraction of obscured QSOs/AGNs appears not constant,
which is higher at low redshift ($z\la 1$) than at high redshifts
($z\ga 1.5$), and the obscured QSOs/AGNs have smaller BHs compared to optically
bright QSOs (e.g.,\ less than a few times $10^8\msun$; see a recent review by
\citealt{Fab03} and references therein).
This observational result contradicts with the expectation from
the simple unification model if the opening angles of the torus are the same for
all QSOs/AGNs, which might suggest that either the opening angle of the torus
is smaller in luminous QSOs/AGNs than in faint QSOs/AGNs
or QSOs/AGNs are obscured only at the early stage of their nuclear
activities \citep{F99}.

The relation between the local BHMF and the QSOLF established in this paper
may provide a way to explore the underlying physics of the
obscured QSOs/AGNs, independent of the X-ray background synthesis model.
With the ongoing and future observations on obscured QSOs/AGNs
(e.g.,\ the X-ray deep surveys by XMM and Chandra), together with
the observations on un-obscured QSOs/AGNs (e.g.,\ by SDSS),
we expect that more accurate constraints on QSO models, the BH growth,
and the physical mechanism of the obscuration can be obtained,
which will improve our understanding of the coevolution of
QSOs and galaxy spheroids.

\subsection{BH mergers after the quenching of the nuclear activity and BH
ejections}
\noindent
The effect of BH mergers on the local BH distribution function, which depends
on the galaxy merger rate and the binary BH evolution process, is still very
uncertain (e.g.,\ \citealt{BBR80,Y02,mm02,vhm02}).  For simplicity, here we only discuss
the effects of mergers after the quenching of the nuclear activity. Since mergers of small BHs
form big BHs, BH mergers may make the BHMF increase at the high-mass end and
decrease at the low-mass end. Thus, the time integral of the QSOLF predicted
from the local BHMF (eq.~\ref{eq:TLlocal}) might be overestimated at the
luminous end and underestimated at the faint end as a result of BH mergers. Indeed, the
tentative result from
the hierarchical galaxy formation scenario shown in Fig.~5 of \citet{KH00} is
that the BHMF at redshift 0 is lower than the BHMFs at higher redshifts
(0.5, 1, 2) by a factor of $<2$ at BH mass $\la 10^{8.5-9}\msun$ and is higher
than those BHMFs at higher redshifts at BH mass $\ga 10^{8.5-9}\msun$. Thus,
at least the models ruled out by applying inequalities (\ref{eq:TMBrelation})
and (\ref{eq:Ropt}) at the luminous end ($M_B\sim -26$) without considering BH
mergers will still be ruled out.

Not all massive BHs may reside in galactic centers (see \S~3.2.1 in
\citealt{YT02}). For example, BHs may be ejected from galactic centers through
either interactions of three or more BHs (e.g.,\ \citealt{V96}) or gravitational
radiation reaction during BH coalescence \citep{R01}, and BHs may also be left
in galactic halos after galaxy mergers if the BH mass ratio of the two merging
galaxies is small enough (e.g., $\la 0.001$; \citealt{Y02}). After considering
the possibility that there might be some BHs that have experienced the QSO
phases not locating in galactic centers, the time integral of the QSOLF
predicted from local galaxies in this paper would increase. Currently, it is
hard to give an estimate on the fraction of the BHs ejected from galactic
centers or left in galactic halos during galaxy mergers by both
theoretical models (also because both the BH and galaxy merger history and
the BBH merger process are still very uncertain) and observations.
\citet{vhm02} study
the assembly and merging history of massive BHs in the hierarchical models of
galaxy formation and argue that the population of BHs wandering in galactic
halos and the intergalactic medium at the present epoch contributes to the
total BH mass density by $\la 10\%$, which (if true) suggests that ignoring BH
ejections would not significantly affect the results of this paper.

A complete and quantitative consideration of BH mergers or ejections
is beyond the scope of this paper.

\section{Other possible applications}\label{sec:application}
\noindent
In this section we discuss two possible applications of
the work established in \S~\ref{sec:relation}, to the study of the
demography of QSOs/AGNs and the demography of the hot stellar components of
normal galaxies at intermediate redshift.

\subsection{Demography of QSOs/AGNs} 
\noindent
Consider such a galactic property ${\cal V}$, which does not significantly
change during the nuclear active phase and after the quenching of the phase
and is closely correlated with the BH mass $M\bhpr$ at present.
Study of the relation between the BH mass and ${\cal V}$ in QSOs/AGNs, as
well as in nearby galaxies, might provide valuable information on QSO
models and BH growth.
Below we show a way to investigate the relation between the BH
mass and ${\cal V}$ in QSOs/AGNs using the work in \S~\ref{sec:relation}.

The posterior distribution of $M\bhpr$ given the luminosity
$L$ of a QSO can be defined as follows:
\be
{\cal P}(M\bhpr|L)\equiv\frac{\int_0^{t_0}\d t\int_0^t\d t_i~{\cal N}(t_i,M\bhpr,L,t)}
{\int_0^\infty\d M\bhpr\int_0^{t_0}\d t\int_0^t\d t_i~{\cal N}(t_i,M\bhpr,L,t)}
=\frac{\tau\life(M\bhpr)P(L|M\bhpr)n_{M\bh}(M\bhpr,t_0)}{\cT\Llocal(L,t_0)},
\label{eq:pmflvb}
\ee
where equations (\ref{eq:inttticalN}) and (\ref{eq:TLlocal}) are used.
Thus, given the local BHMF and the luminosity evolution of
individual QSOs ${\cal L}(M\bhpr,\tau)$, which can be used to obtain
$\tau\life(M\bhpr)P(L|M\bhpr)$,
we may use equation (\ref{eq:pmflvb}) to get ${\cal P}(M\bhpr|L)$ and further
use the local $M\bhpr-{\cal V}$ relation and get the distribution of
${\cal V}$ at a given $L$ or $M\bh$ in QSOs.
Comparison of future observation results (see also current observational
results on the BH mass-velocity dispersion relation in QSOs by
\citealt{Shields03}) with
the expected relations between the galactic parameter ${\cal V}$ and the
luminosity/BH mass of QSOs may further constrain the QSO luminosity evolution.

In a separate paper \citep{YL03} we will explore the
nuclear luminosity/BH mass and velocity dispersion relation in QSOs/AGNs
in the way described above, assuming that the velocity dispersion
of the hot stellar components of galaxies does not significantly change 
during the nuclear active phase and after the quenching of the phase.

\subsection{The distribution of velocity dispersions in elliptical
galaxies and bulges of S0/spiral galaxies at intermediate redshift}
\noindent
Study of the demography of galaxies at all redshifts may help us understand
the formation and evolution of galaxies.
The past decade has seen dramatic expansion of the knowledge on
intermediate- and high-redshift galaxies in both observations and theories,
as well as on local galaxies,
such as the observations of the Lyman-break galaxies at redshift $z\sim 2-4$
(e.g.,\ \citealt{Adelberger03}),
the semi-analytic hierarchical galaxy formation models to explain
and predict the observational properties of both early-type and late-type
galaxies (e.g.,\ \citealt{WF91,KWG93,Cole94,Cole00}), etc.
Below using the work in \S~\ref{sec:relation},
we point out a simple way to estimate the velocity dispersion distribution in
elliptical galaxies and the bulge components of S0/spiral galaxies at
intermediate redshift ($1\la z\la 2.5$ here), which are still currently
poorly known and are expensive and difficult to measure by observations.

We define $n_{M\bh}(M\bhpr,t)$ as the BHMF of dead QSOs at cosmic time $t$ so
that $n_{M\bh}(M\bhpr,t)\d M\bhpr$ represents the comoving number density
of the BHs whose nuclear activities had been quenched before time $t$ and
the mass of these BHs is in the range $M\bhpr\rightarrow M\bhpr+\d M\bhpr$ from 
the quenching time to the present time $t_0$.
Similarly as the derivation to get equations (\ref{eq:TLlocal}) and
(\ref{eq:Trelation}), we have 
\be
\int^t_0\Psi_L(L,t)dt=\int^\infty_0 \d M\bhpr \tau(M\bhpr)P(L|M\bhpr)
n_{M\bh}(M\bhpr,t),
\label{eq:lfatbigz}
\ee
as long as the condition
\be
\int^t_0\Psi_L(L,t)dt\gg \int^t_{t-\tau\life}\Psi_L(L,t)dt
\label{eq:lfcond}
\ee
is satisfied.
Condition (\ref{eq:lfcond}) represents that most QSOs contributing to the
integral of the QSOLF over the cosmic time $0-t$ have become quiescent
(note the similarity to the necessary condition to hold eq.~\ref{eq:TLlocal}
in \S~\ref{sec:relation} that most or all QSOs are quiescent at present time
$t_0$).
Using the observed QSOLFs shown in \S~\ref{sec:QSOLF}, we find that condition
(\ref{eq:lfcond}) can be satisfied at redshift $z\la2.5$
if $\tau\life\la 4\tau\Sp(\epsilon=0.1)\simeq 2\times 10^8\yr$.

As shown in \S~\ref{sec:constraint}, the comparison between relation
(\ref{eq:Trelation}) and observations may provide constraints on the QSO
luminosity evolution or $\tau(M\bhpr)P(L|M\bhpr)$.
Given $\tau(M\bhpr)P(L|M\bhpr)$ and the QSOLF,
$n_{M\bh}(M\bhpr,t)$ can be uniquely determined from equation
(\ref{eq:lfatbigz}) (which is just a typical inverse problem).
According to Bayes's theorem, the distribution of the velocity dispersion
of the hot stellar components of local  galaxies given a BH mass $M\bhpr$ is given by:
\be
P(\sigma|M\bhpr)=\frac{P(M\bhpr|\sigma)n_\sigma(\sigma,t_0)}
{n_{M\bh}(M\bhpr,t_0)}.
\label{eq:probmfsig}
\ee
We assume that the formation and the significant part of the evolution of
bulges are simultaneous with the significant evolution (and/or formation) of
their central BHs and that their velocity dispersions (and BH masses) do not
significantly change after the quenching of the nuclei activity.
Thus, there will be little evolution of the BH mass and velocity dispersion
relation in those galaxies after they have experienced QSO phases, and the
significant part of the evolution of the BH mass-velocity dispersion
relation is recorded only in QSOs/AGNs.
With $P(\sigma|M\bhpr)$ obtained in equation (67),
the velocity dispersion distribution of the hot stellar components of
galaxies at low and intermediate redshift ($z\la 2.5$) can be given by
\be
n_\sigma(\sigma,t)=\int^{\infty}_0 \d M\bhpr P(\sigma|M\bhpr)
n_{M\bh}(M\bhpr,t).
\label{eq:nsigmat}
\ee
Note that according to the assumptions above, for normal galaxies, the hot
stellar components exist only in those galaxies containing massive BHs (BH
ejections are ignored here), and those galaxies that have experienced QSO
phases and contain BHs must have hot stellar components, which of course
should be tested by future observations. The methods above would
still be applicable even if not all of the bulges or galaxies containing BHs follow these assumptions, but
as long as most of them do.
In addition, if the formation of the hot stellar components occurs before
the formation of central BHs or the QSO phase (that is, some hot stellar
components may not contain BHs), the velocity dispersion distribution of the
galaxies obtained by the method in this section will at least give the lower
limit to their realistic distribution.

\section{Conclusions}\label{sec:conclusion}
\noindent
Assuming that each massive BH in nearby galactic centers has
experienced the QSO phase and becomes quiescent at present,
we have established a relation between the QSOLF and the local BHMF
by studying the continuity equation for the BH mass and nuclear luminosity
distribution and ignoring BH mergers.
This relation compares the time integral of the QSOLF and that
inferred from the local BHMF and only incorporates the luminosity evolution
of individual QSOs. The triggering history of the accretion onto seed BHs is
(implicitly) considered in the continuity equation but is circumvented in
the relation between the QSOLF and the local BHMF.
For comparison, the old relations between QSOs and local BHs on the
total/partial BH mass densities (see eqs.~\ref{eq:soltan} and \ref{eq:YT};
\citealt{S82,YT02})
include the effect of BH mergers, but the seed BH mass is ignored.
The relation on the total BH mass density (eq.~\ref{eq:soltan}; \citealt{S82})
is unrelated with the luminosity evolution of individual QSOs, and the relation
on the partial BH mass density (eq.~\ref{eq:YT}; \citealt{YT02}) assumes that
the luminosity of QSOs is only an increasing function of their central BH mass
(e.g.,\ the Eddington luminosity in the calculation in \citealt{YT02}).
The new relation on the time integral of the QSOLF in this paper can be used
to explore the luminosity evolution of individual QSOs (see \S~\ref{sec:intro}
and \ref{sec:degeneracy}).

By applying observations into the relation established in this paper and
assuming that the nuclear luminosity evolution includes two phases (first
increasing at the Eddington luminosity with the BH growth and then declining),
we find that the time integral of the QSOLF is generally consistent with that
inferred from local BHs and obtain the following observational constraints on
the QSO luminosity evolution and BH growth:
(i) The QSO mass-to-energy efficiency $\epsilon$ should be $\ga 0.1$
(see Fig.~\ref{fig:eff}).
(ii) The lifetime (defined directly through the luminosity evolution of 
individual QSOs) should be longer
than $\tau\Sp$($\simeq 5\times 10^7\yr$) if $\epsilon=0.1$
and $0.2\tau\Sp$($\simeq 4\times 10^7\yr$) if $\epsilon=0.31$
(see Figs.~\ref{fig:model1sta} and \ref{fig:eff}).
The characteristic declining timescale in the second phase should be
significantly shorter than $\tau\Sp$, and BH growth should not be dominated
by the second phase (when QSOs are accreting at sub-Eddington luminosities;
see Fig.~\ref{fig:model3}).
(iii) The upper limit of the ratio of obscured QSOs/AGNs to optically bright
QSOs is provided, which should be not larger than 7 at $M_B\sim -23$ and
3 at $M_B\sim -26$ if $\epsilon=0.31$ and not larger than 1 at $M_B\sim -23$
and negligible at $M_B\sim -26$ if $\epsilon=0.1$
(see Figs.~\ref{fig:model1sta} and \ref{fig:eff}).
(iv) It is unlikely that most QSOs are accreting at
super-Eddington luminosities (see Fig.~\ref{fig:supedd}).
The constraints above are obtained by assuming that the two accretion phases appear
only once in the luminosity evolution, although this assumption is not required in
the relation established in this paper.
The possibility of more complicated accretion patterns
deserves further investigation.

We find that if the QSO lifetime is longer than a certain value
(e.g.,\ $\sim 4\tau\Sp$; see Fig.~\ref{fig:model1sta}), the time integral
of the nuclear LF inferred from local BHs becomes insensitive to the
value of the QSO lifetime, and thus it is difficult to provide an accurate
estimate on the QSO lifetime unless observations extend to fainter
luminosities or precise measurements of the QSOLF and local BHMF are available
(e.g.,\ with error much less than 10\%).
We also point out that this difficulty would also
exist in many other methods to estimate the QSO lifetime by using the QSOLF,
as a result of the sharp decrease of the QSOLF at the bright end and the limited
luminosity range in observations.

We show the importance of accurately measuring the intrinsic scatter in
the relation between the BH mass and velocity dispersion of local galaxies 
and the scatter in the distribution of bolometric corrections of QSOs to 
precise understanding of the physics behind the QSO phenomenon and BH growth.
Both of the scatters affect the shape and values of the time-integral of 
the nuclear LF, especially at the bright end.

With the upcoming more precise measurement on QSOs (including both
unobscured and obscured AGNs) and the demography of local BHs and galaxies
(by SDSS, Chandra, XMM, etc.), 
the method presented in this study would help to further explore the nuclear
activity triggering and quenching mechanisms, obscuration of QSOs/AGNs, the
demography of QSOs/AGNs, and the demography of normal galaxies at intermediate
redshift and finally understand the physics behind the
QSO phenomenon and the formation and evolution of galaxies.

\acknowledgments
We thank Ravi Sheth for helpful communication on the velocity
dispersion distribution of nearby galaxies. We thank Norm Murray for helpful
discussions. We thank the referee and Scott Tremaine for thoughtful comments.

\end{document}